\documentclass[journal]{IEEEtran}
\usepackage{cite}

\ifCLASSINFOpdf
  \usepackage[pdftex]{graphicx}
\else
\fi
\usepackage{amssymb}
\usepackage{amsthm}
\usepackage{amsmath}

\usepackage{ulem}
\usepackage{xspace}

\usepackage[colorlinks]{hyperref}
\hypersetup{
    colorlinks=true,
    linkcolor=blue,
    filecolor=magenta,      
    urlcolor=cyan,
    pdftitle={Sharelatex Example},
    pdfpagemode=FullScreen,
}
 
\graphicspath{ {./pics/} }

\newcommand{\Ti}{\ensuremath{T_{i}}\xspace}
\newcommand{\Te}{\ensuremath{T_{e}}\xspace}
\newcommand{\Tig}{\ensuremath{T^{g}_{i}}\xspace}
\newcommand{\Teg}{\ensuremath{T^{g}_{e}}\xspace}
\newcommand{\Tim}{\ensuremath{T^{m}_{i}}\xspace}
\newcommand{\Tem}{\ensuremath{T^{m}_{e}}\xspace}

\newcommand{\ig}{\ensuremath{i_{g}}\xspace}
\newcommand{\im}{\ensuremath{i_{m}}\xspace}
\newcommand{\ic}{\ensuremath{i_{c}}\xspace}
\newcommand{\ip}{\ensuremath{i_{p}}\xspace}
\newcommand{\ifi}{\ensuremath{i_{f}}\xspace}
\newcommand{\ia}{\ensuremath{i_{a}}\xspace}

\newcommand{\Ed}{\ensuremath{E_{d}}\xspace}
\newcommand{\Et}{\ensuremath{E_{t}}\xspace}

\newcommand{\dV}{\ensuremath{\Delta V}\xspace}

\newcommand{\Fe}{$^{55}\mathrm{Fe}$\xspace}

\newcommand{\pAs}{$\mathrm{picoammeters}$\xspace}

\newcommand{\anode}{\textrm{anode}\xspace}
\newcommand{\field}{\textrm{field}\xspace}
\newcommand{\pad}{\textrm{pad}\xspace}
\newcommand{\mesh}{\textrm{mesh}\xspace}
\newcommand{\cath}{\textrm{cathode}\xspace}

\newcommand{\BPG}{\textrm{BPG}\xspace}
\newcommand{\IBF}{\textrm{IBF}\xspace}
\newcommand{\TPC}{\textrm{TPC}\xspace}
\newcommand{\TPCs}{\textrm{TPCs}\xspace}
\newcommand{\GEM}{\textrm{GEM}\xspace}
\newcommand{\FoM}{\textit{FoM}\xspace}

\begin{document}
%
\title{Measurement of the ion blocking by the passive bi-polar grid}
%
%
%


\author{\IEEEauthorblockN{E. Shulga,
V. Zakharov, 
P. Garg,
T. Hemmick, and
A. Milov}
\thanks{This research is supported by grant number 2016240 from the United States-Israel Binational Science Foundation (BSF),  Jerusalem, Israel.}
\thanks{V. Zakharov, P. Garg and T. Hemmick are with Department of Physics and Astronomy, Stony Brook University, Stony Brook, NY, USA.}
\thanks{E. Shulga and A. Milov are with Department of Particle Physics and Astrophysics, Weizmann Institute of Science, Rehovot, Israel.}
\thanks{Manuscript received July 27, 2020; revised October 13, 2020.
Corresponding author: E. Shulga (email: \href{mailto:evgeny.shulga@weizmann.ac.il}{evgeny.shulga@weizmann.ac.il}).}
}



\date{October 2020}

\maketitle

\begin{abstract}
The ion backflow is the main limiting factor for operating time projection chambers at high event rates. A significant effort is invested by many experimental groups to solve this problem. This paper explores a solution based on operating a passive bi-polar wire grid. In the presence of the magnetic field, the grid more effectively attenuates the ion current than the electron current going through it. Transparencies of the grid to electrons and ions are measured for different gas mixtures and magnitudes of the magnetic field. The results suggest that in a sufficiently strong magnetic field, the bi-polar wire grid can be used as an effective and independent device to suppress the ion backflow in time projection chambers.
\end{abstract}


%
\IEEEpeerreviewmaketitle

\section{Introduction}
\label{sec:intro} 

The time projection chambers (\TPCs) are introduced by David Nygren~\cite{Nygren:1974wta} in 1974 and have been successfully used in different particle physics experiments~\cite{Aihara:4332223,Kamae:1986jd,BRAND1989567,ATWOOD1991446,FUCHS1995394,Anderson:2003ur,ALME2010316}. \TPCs have a number of features that make them an attractive technological choice for detectors in high-energy and nuclear physics experiments. 

Due to their excellent capability to reconstruct 3D topology of charged particles produced in interactions \TPCs are widely used in experiments where the measurement of a multi-particle final state is required. Being operated in an external magnetic field, \TPCs provide high precision momentum measurement of the tracks, down to very low magnitudes. Sampling energy deposition in the gas working volume gives \TPCs the particle identification capabilities. The use of the gas as a working medium makes \TPCs a low radiation length detectors that are easily combined with detectors based on other technologies as it is required in most modern experiments. Last but not least, \TPCs are relatively inexpensive devices. A combination of these features makes \TPC a widely used detector technology after more than 45 years since it was introduced.

Together with the advantages, the main setback of \TPCs is the low data taking rate which is a severe constraint on the use of \TPCs in modern experiments requiring high data-taking rates. Among several factors that affect the rates the most difficult to overcome is the space charge that builds up in the \TPC volume and distorts the drift of the primary ionization. Charges in the \TPC volume are carried by slow-moving ions produced in the readout elements of the \TPC. This is known as the positive ion backflow (\IBF) problem.

To address the \IBF problem the first \TPC built in 1984~\cite{Aihara:4332223} used a plane of wires called the bipolar gating grid (\BPG) separating \TPC readout elements from the drift volume. Applying positive and negative bias voltages to odd an even wires of the grid stops the ion and electron flow through the \BPG. \TPCs developed in recent years~\cite{Ball:2012xh,Abelevetal:2014cna,COLAS2004226} adopt the concept of amplification element being also the \IBF-stopper. Multiple-layer micropattern detectors used as amplification elements are capable of trapping ions between their layers~\cite{SAULI2006269,1312013,BONDAR2003325,BLATT2006155,Chechik:2003hx,5658080,Bohmer:2012wd,Aiola:2016rld,WANG2019410}. Nevertheless, most of the large \TPCs built by the present time rely on the \BPG to suppress the \IBF~\cite{Hilke_2010}. 

A \BPG can be operated in synchronous and passive modes~\cite{AMENDOLIA198547,AMENDOLIA1986403}. The former implies that the voltage bias on the wires is synchronized with an external trigger. The duration and the frequency of pulses ensure that all ions are collected on the \BPG. It also results in stopping the electrons going through the \BPG, producing a dead time in the system. In the presence of the magnetic field, the voltages required to stop electrons are higher than those that are required to stop the ions, which allows the \BPG to retain some electron transparency when the ion current is fully shut. It opens the possibility to operate the \BPG in passive mode with constant biases on odd an even wires. Achieving high data-taking rates in a \TPC operated in a passive mode is much easier. All \TPCs built by the large particle experiments up to the present time used the \BPG in a synchronous mode, although passive mode was also considered for the detectors in the magnetic fields above 1~T~\cite{Anderson:2003ur,ALME2010316}. 

The principle of the \BPG operation in a passive mode is based on the effect that in the presence of the magnetic and electric fields the direction of the electron drift has a component along the vector product of the two fields. The electron drift in this direction is described with the Lorentz angle that is explained in many works~\cite{Hilke_2010,Kent:1984ha,Sauli:1992bu,Blum:1105920} and therefore is not elaborated here. This paper provides a detailed study of the \BPG transparency for electrons and ions in different gas mixtures in the presence of the magnetic field. The results show that the \BPG operated in a passive mode can be used as an effective element to suppress the \IBF in the \TPCs operated in a strong magnetic field, for example in the sPHENIX \TPC~\cite{Adare:2015kwa,sPHENIX:2015irh}.
\section{Measurements}
\label{sec:mesurements}
\subsection{The setup}
\label{sec:setup}
\begin{figure*}[ht]
   \centering
   \includegraphics*[width=\textwidth]{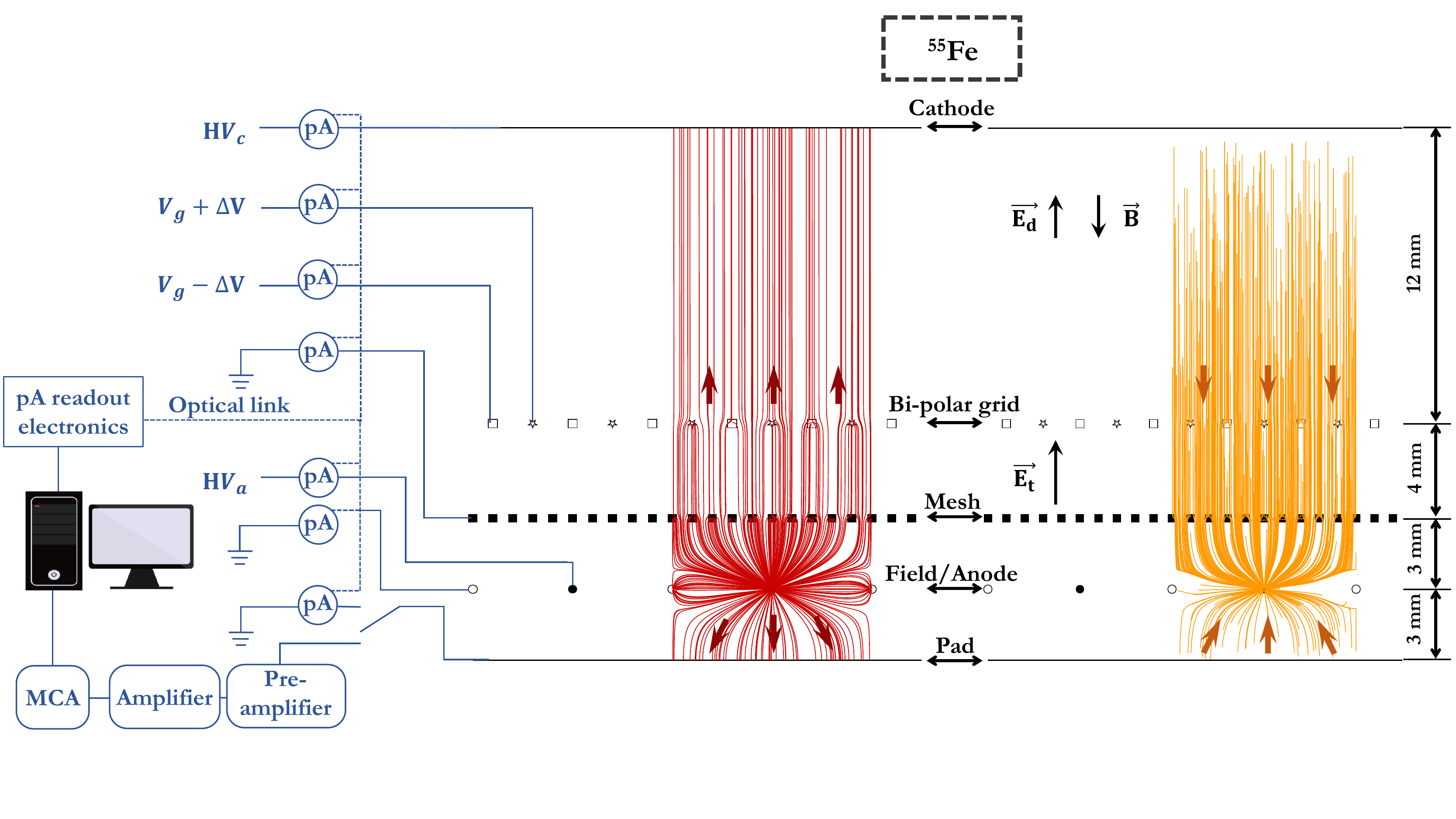}
   \caption{Schematic view of the experimental setup. Colored (online) lines represent ion (left) and electron (right) drift trajectories, respectively.}
   \label{fig:setup}
\end{figure*}

The setup built at the Weizmann Institute consists of the \BPG sandwiched between the ion generating plane and the ion receiving cathode immersed into a gas volume. A schematic view of the setup is shown in Fig.~\ref{fig:setup}. 

All frames used in the setup have the working area of $27\times27$ mm$^{2}$. The primary ionization is produced by the \Fe source positioned above the gas volume. At the time of measurement, the source intensity was approximately 3.5~mCi. The amount of gamma radiation illuminating the detector volume is controlled by a collimator. 

X-rays from the source enter the gas volume through a \cath electrode built of a thin \GEM electrically connected on both sides and illuminates the entire detector. A volume between the \cath and the \BPG has a vertical dimension of 12~mm. Photons converting in this volume produce primary ionization that drifts towards the \BPG. The \BPG is built of 50~$\mu$m wires spaced by 1~mm. Odd and even wires of the \BPG are set to a voltage of $V_{g}\pm\dV$ fed on the opposite sides of the frame. Thus the adjacent wires have a voltage difference of $2\dV$. The \mesh electrode is located 4~mm below the \BPG. It is made of the stainless steel mesh with $0.5\times0.5$~mm$^{2}$ cells and a wire diameter of 50~$\mu$m, providing $\sim$80\% optical transparency. Electrons passing the \BPG and those that are coming from conversions inside the 4~mm space, enter the \mesh electrode. A collimator (not shown in the figure) that immediately follows the \mesh is made of thin dielectric material and limits the working area of the detector to a circle of 20~mm in diameter. The collimator eliminates the edge effects and increases lateral uniformity of the ionization flux.

The wire plane located 3~mm below the \mesh is made of 50~$\mu$m Cu/Rh wires spaced by 2.5~mm. Voltages are applied to the wires on opposite sides of the frame. Field wires are grounded and \anode wires are set to 1.7--2.1~kV to provide the desired gas gain depending on the gas mixture. To reduce parasitic currents flowing between \field and \anode wires the grooves are made in the FR4 material of the frames holding the wires. Wires and grooves are covered with epoxy. During the assembly, the \anode and \field wires on the wire plane are directed orthogonal to the \BPG wires. This plane is located 3~mm above the \pad plane that is grounded. Electrons from the conversion of the \Fe photons that occur in this volume reach the \anode wires without passing through any other element in the setup.

The setup is shown in Fig.~\ref{fig:setup}. It is assembled in a $15\times15\times3$ cm$^{3}$ dielectric box and covered with a copper foil on the outside for electrical grounding. The vertical size of the box is constrained by the dimension of the magnet bore in which the dipole field up to 1.2~T is generated by a magnet produced by Danphysik GGG. The field is controlled by Group3 DTM-151 tesla meter with MPT-141 probe providing 0.012\% accuracy. 

The HV is supplied by CAEN N471 and Lambda Z$^{+}$ 320 power supplies through the low-pass filters with $RC\approx2$~s. All conductive elements in the setup are read out by the floating picoammeters connected to the computer via optical links. Picoammeters are produced by PicoLogic J.D.O.O. in Zagreb~\cite{UTROBICIC201521}. In the working regime, the parasitic currents in the measured channels averaged over 1~s, are in 5--80~pA range at the highest \anode voltages. Signals from the \pad electrode can be switched to the charge measuring channel consisting of the Ortec charge sensitive preamplifier 142 IH followed by a shaping amplifier Ortec 672 and read out by the Ortec multichannel analyzer (MCA), Ametek Easy-MCA 2000.

The gas mixtures are prepared in a gas mixing station using calibrated mass flow controllers (Aalborg GFCS). Flow controllers are calibrated by the water displacement method after each change of mixed gases. The accuracy of the quenching fraction in the gas mixture is $\pm$3.5\% for (90:10) gas mixtures and $\pm$1.5\% for (50:50) gas mixtures. Gas flow is set to provide detector volume exchange every 10 minutes preventing possible outgassing from the structural elements of the setup into the working gas atmosphere. The gas used in the measurement comes in the bottles with purity $>99.99$\% and is not recirculated during the measurements.

\subsection{Definitions of transparency}
\label{sec:method}

The measurements are carried out at a sufficiently high gain on \anode wires such that the contribution of the primary ionization to the currents ($i$) can be neglected. Then 
\begin{equation}
-\ia=\ic+\ig+\im+\ifi+\ip 
\label{eq:currents}
\end{equation}
is fulfilled with the accuracy of picoammeters. Currents in the equation correspond to \cath~(\ic), \BPG~(\ig), \mesh~(\im), \field~(\ifi), \pad~(\ip), \anode~(\ia) electrodes respectively, as shown in Fig.~\ref{fig:setup}.

The \BPG is considered here as a standalone element in an arbitrary \TPC detector. For electrons and ions traversing the \BPG, its impact can be characterized by transparency parameters denoted as \Teg and \Tig respectively. From the setup shown in Fig.~\ref{fig:setup} and Eq.~\eqref{eq:currents}.
\begin{eqnarray}
\Tig=\frac{\ic}{-\ia-\ip-\im-\ifi}.
\label{eqn:tig}
\end{eqnarray}
That is the ion current reaching the \cath above the \BPG divided by the ion current that flows into the \BPG, {\it i.e.} the ion current emerging from \anode wires less the currents in \pad, \field, and \mesh electrodes. Analogously, one can also define the ion transparency of the \mesh electrode.
\begin{eqnarray}
\Tim=\frac{\ic+\ig}{-\ia-\ip-\ifi}.
\label{eqn:tim}
\end{eqnarray}

\Teg cannot be defined as a ratio of currents, because the electron components in all currents, except in \ia, are negligibly small. The value of \ia depends on the amount of initial ionization in all parts of the setup. Neglecting electron attachment, electrons from photon conversions in the gas arrive at anode wires unless they are captured by the \BPG or the \mesh. By changing \dV on \BPG one can suppress or fully block electrons coming from the detector volume above the \BPG. This consideration leads to Eq.~\eqref{eqn:shape} which is the ratio of the anode current, coming from everywhere in the setup, to the current that is coming from below the \BPG. \Teg can be deduced from the shape of \ia measured as a function of \dV.
\begin{eqnarray}
\frac{\ia(\dV)}{\ia(\Ed=0)} = 1 + K\Tem\Teg(\dV),
\label{eqn:shape}
\end{eqnarray}
The constant coefficient of $K$ in the equation is a relative amount of primary ionization that is generated above and below the \BPG and amplified on the \anode wires. The $\Tem$ is the \mesh transparency to electrons. To first order, these coefficients do not depend on \dV since the electric fields around the \mesh are not affected by the voltages between the \BPG wires, see Fig.~\ref{fig:setup}.

The \anode current on the left-hand side of the equation is divided by the current, measured when electrons from above are not flowing to the \BPG, $\ia(\Ed=0)$. It is experimentally proven that the same current is measured in the \anode wires when the \BPG is fully closed to primary electrons. In this case, the reduced \ia is equal to unity. Thus, the shape of the $\Teg$ dependence can be extracted from measuring $\ia(\dV)$ and using Eq.~\eqref{eqn:shape}. Determining the absolute normalization of \Teg from Eq.~\eqref{eqn:shape} requires precise knowledge of the coefficient $K$, which in turn depends on the geometry of all elements in the setup. Instead, for the final results, the absolute normalization of \Teg was worked out as follows.

Charge distributions from the \Fe ionization source are measured in the \pad electrode for three different cases: ionization that is coming to the \anode directly; ionization reaching the \anode through the \mesh; ionization reaching \anode through \mesh and \BPG. These distributions are obtained by setting the drift field (\Ed) between the \cath and the \BPG and the transfer field (\Et) between the \BPG and the \mesh to their nominal values or zero. The three distributions are shown with lines in the left panel of Fig.~\ref{fig:charges}, the corresponding field settings are mentioned in the figure legend.
\begin{figure*}[htb!]
\centering
\includegraphics[width=0.49\textwidth]{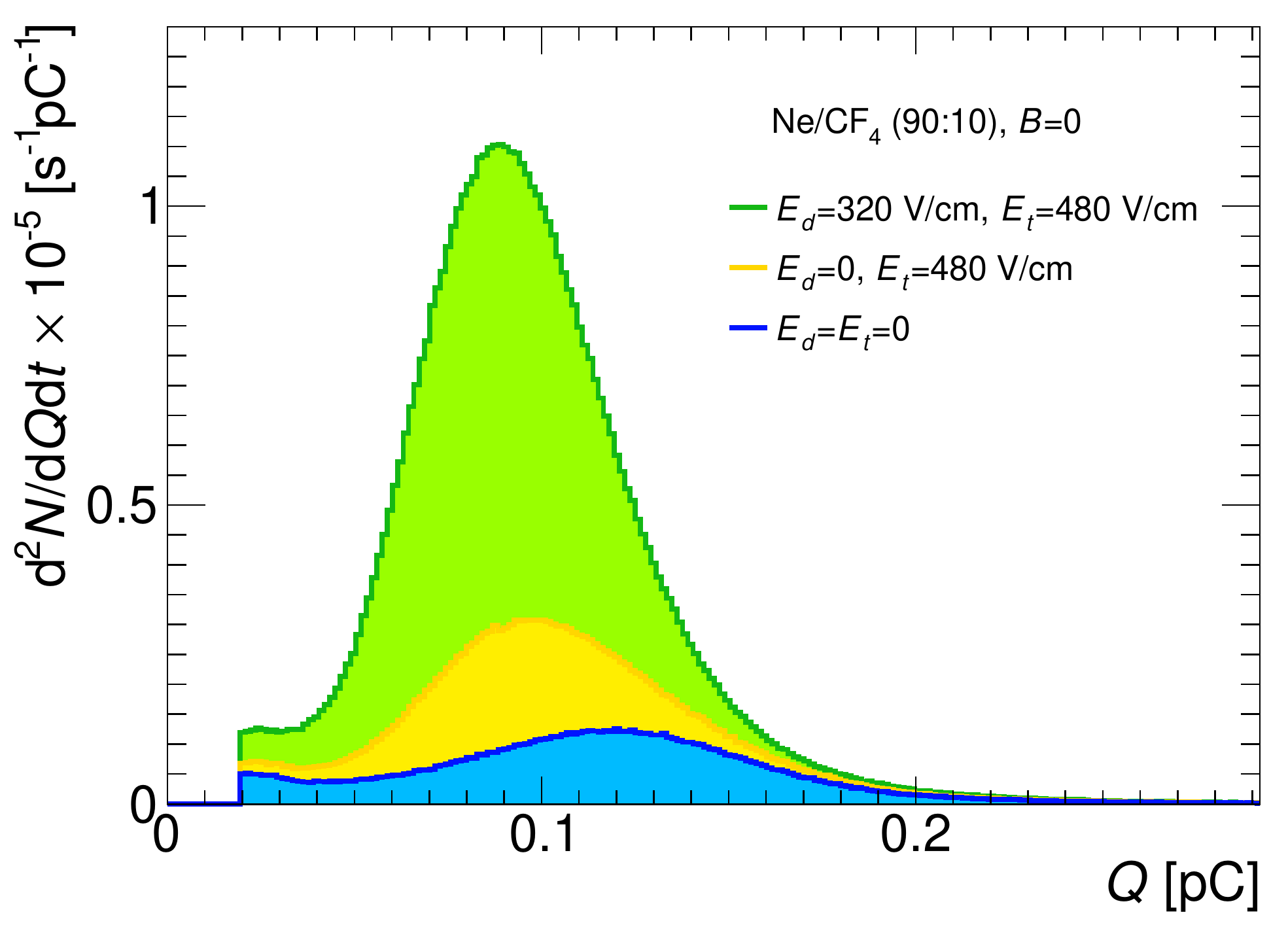}
\includegraphics[width=0.49\textwidth]{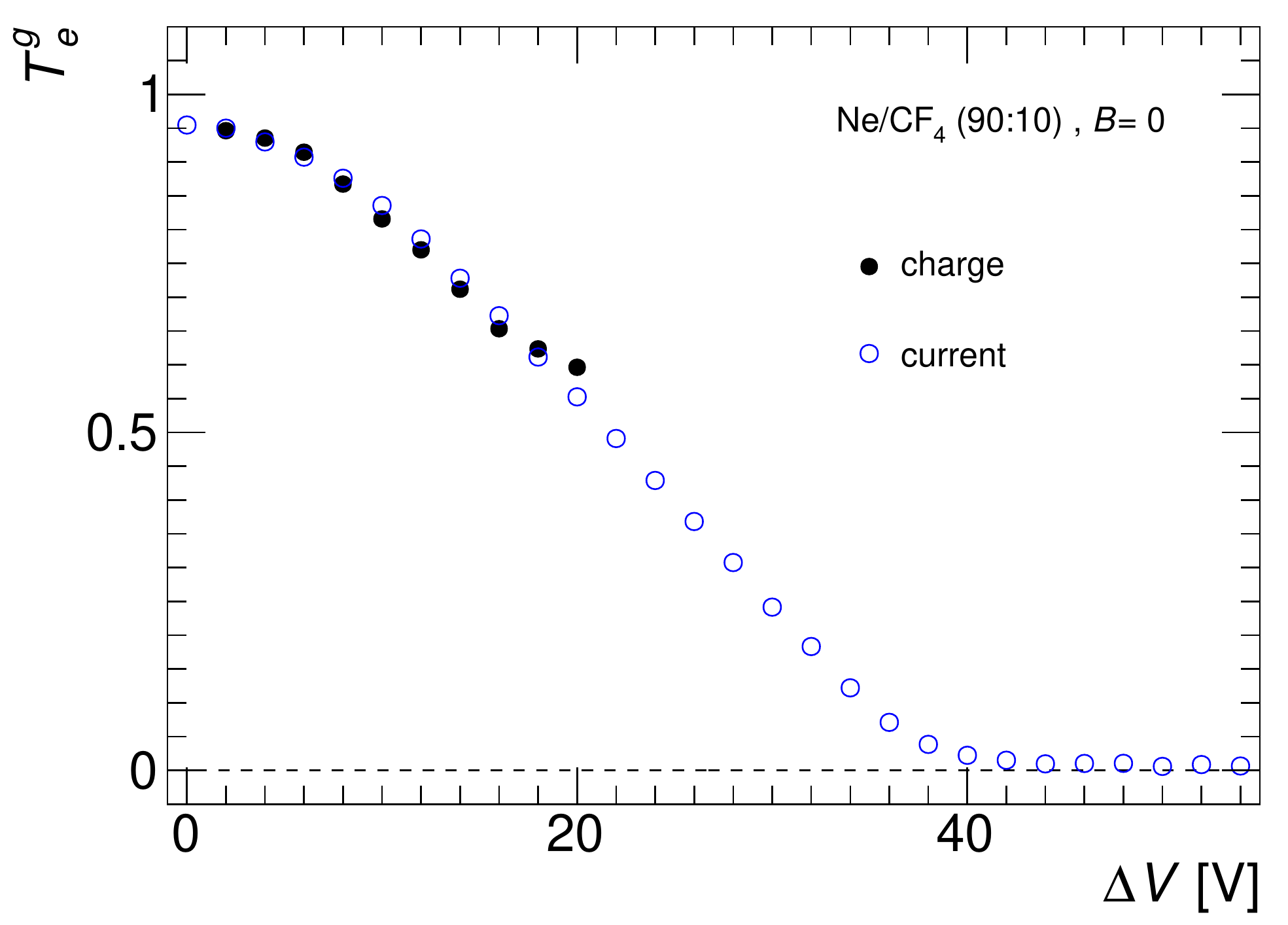}
\caption{Left: Charge distributions measured in the \pad electrode at different field configurations. Right: \Teg as a function of \dV calculated from the currents and the charge measurements. The magnitude of the current measurement at $\dV=0$ is set to the value of the charge measurement and therefore is not shown.}
\label{fig:charges}
\end{figure*}
Primary ionization from the photons converting above the \BPG drifts through the \BPG and the \mesh and therefore is attenuated by the factor (\Teg\Tem). Primary ionization from the photons converting between the \BPG and the \mesh is attenuated by the factor of \Tem only. Therefore by subtracting distributions one can compare two distributions before and after the \BPG. The ratio of the mean values of theses two distributions after extrapolating them to zero provides the absolute normalization of \Teg. The right panel of Fig.~\ref{fig:charges} shows the result of this measurement with filled markers. The results are compared to the current measurement based on Eq.~\eqref{eqn:shape} and shown with open markers. All points in the current measurement are multiplied by the same factor such that the first point, at $\dV=0$, gets the value of the charge measurement. After this both curves agree in the region $\dV<20$~V. Above 20~V the charge measurement becomes difficult because peaks shown in the left panel of the figure disappear and distributions shift close to zero values. Results presented in this paper are based on the measurement deduced from the currents and normalized from the charges.

\subsection{Source intensity}
\label{sec:spece-charge}

The amount of ionization let into the system by the collimator is chosen as a compromise. Lowering currents to a few pA level requires the extension of the measurement cycle to hours, and makes smaller effects enter the consideration such as control over detector stability, better knowledge of the baseline values, additional control over low-frequency micro discharges, etc. Raising the currents results in the build-up of the space charge in the setup that alters electric fields around the \BPG. Although the space charge problem is typically associated with much larger detectors, the setup is shown in Fig.~\ref{fig:setup} with the working area of approximately 3~cm$^2$ and relatively short gaps operates at much higher current densities than most of the larger detectors.

The space charge effects are studied with the \Tig curve measured with different currents. It is done by attenuating \Fe source with the collimator and by lowering the gas gain. The results are shown in Fig.~\ref{fig:current_test}. Current ranges are characterized by the \cath current $\ic(\dV=0)$, typically the largest current used to produce the corresponding curve.
\begin{figure}[htb]
   \centering
   \includegraphics*[width=0.49\textwidth]{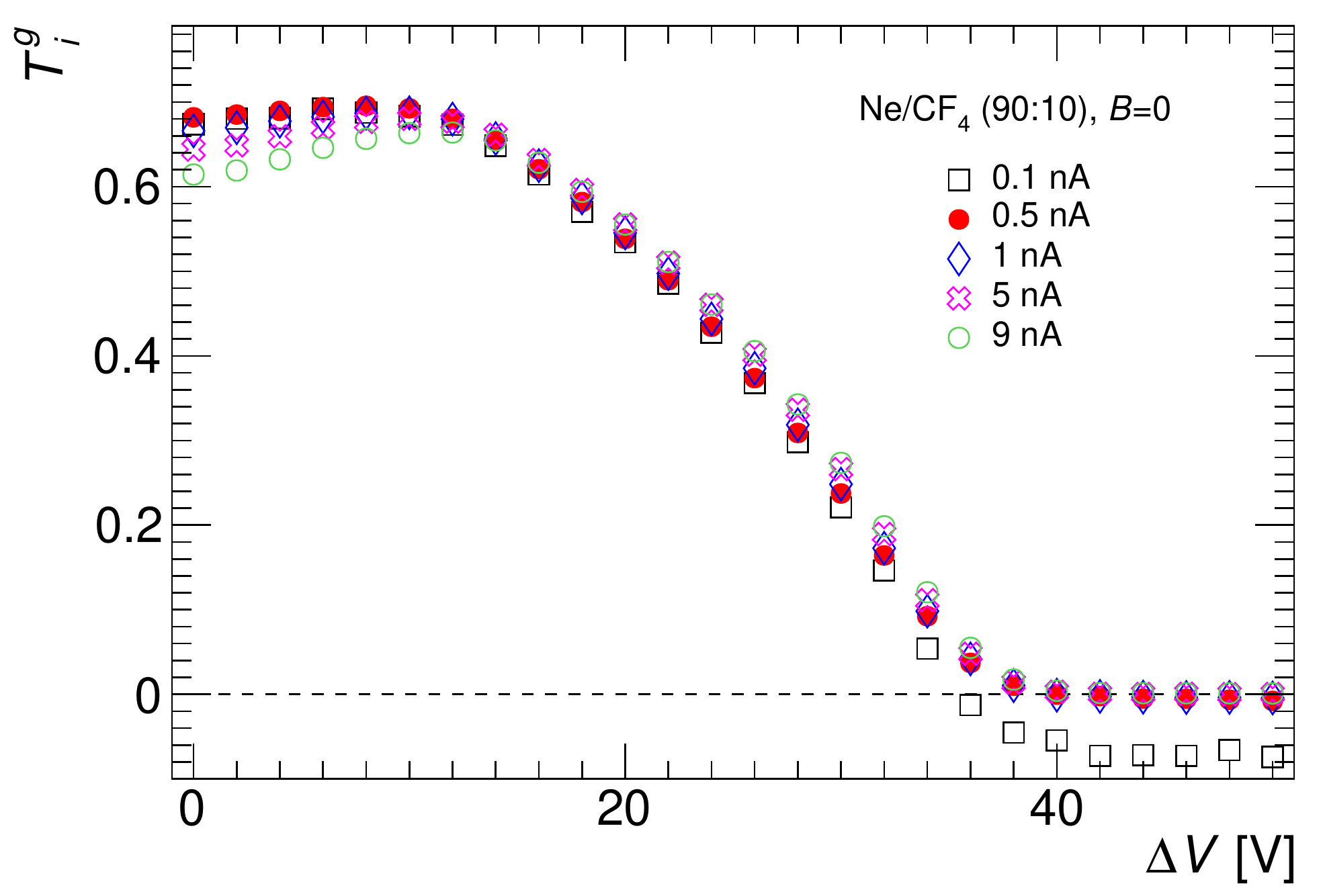}
   \caption{\Tig as a function of \dV measured using different current ranges indicated with \ic(\dV$=0$) in the legend.}
   \label{fig:current_test}
\end{figure}
The study shows that the measurements made with currents below 500~pA are hard to reproduce and are considered unstable, whereas in the range above 1~nA the space-charge effects start to develop at $\dV=0$. Thus for the results presented in this paper the currents in the \cath are kept below 1~nA to avoid space-charge effects and prevent zero-current miss-measurement.

\subsection{Measurement procedure}
\label{sec:proc}
\subsubsection{General settings}
All measurements follow the standard procedure. The gas flow is set to approximately 30~cm$^{3}$/min and the detector is flushed over 1~h, corresponding to more than 10 detector and tubing volumes exchanges. The magnetic field is set to the desired value. Voltages are set on \anode, \cath, and both types of the \BPG wires. Pad, \field, and \mesh electrodes remain grounded. The gas gain is set to the nominal value of 3500 measured by the position of ionization peak in the \pad. The collimator is adjusted to produce the \ic current from the \Fe source close to 1~nA. The detector is operated in this stage for approximately 30~min after which the settings are additionally adjusted if needed.

The following three measurements are performed in each gas mixture at each magnetic field magnitude and combined to produce the final results reported in this paper.

\subsubsection{Transfer field scan}
\label{sec:et_measurement}
    
The \BPG transparency to ions and electrons strongly depends on the magnitudes of \Ed and \Et, which choice is closely related to properties of gas and many other considerations~\cite{ALME2010316,603731}, including the Lorentz angle. Since optimization of the gas mixture and \Ed is not feasible in the scope of this paper, to make comparative studies of the \BPG performance, different gases are measured in the same field configuration, called "main" in which \Ed is kept at a constant value of 320~V/cm in all measurements. To find dependence on \Et the voltages on \cath and \BPG are set to values that provide $\Ed=320$~V/cm and $\Et=0.5\Ed =160$~V/cm. The two voltages are then increased in steps such that \Et is incremented by $0.125\Ed$ until it reaches $2.5\Ed$. After each voltage change, the detector is operated for a waiting time of 5~min without any change and then the measurement is taken averaged over 1~min.

Results of the \Et scan are shown in Fig.~\ref{fig:edscan}.
\begin{figure}[htb]
   \centering
   \includegraphics*[width=0.49\textwidth]{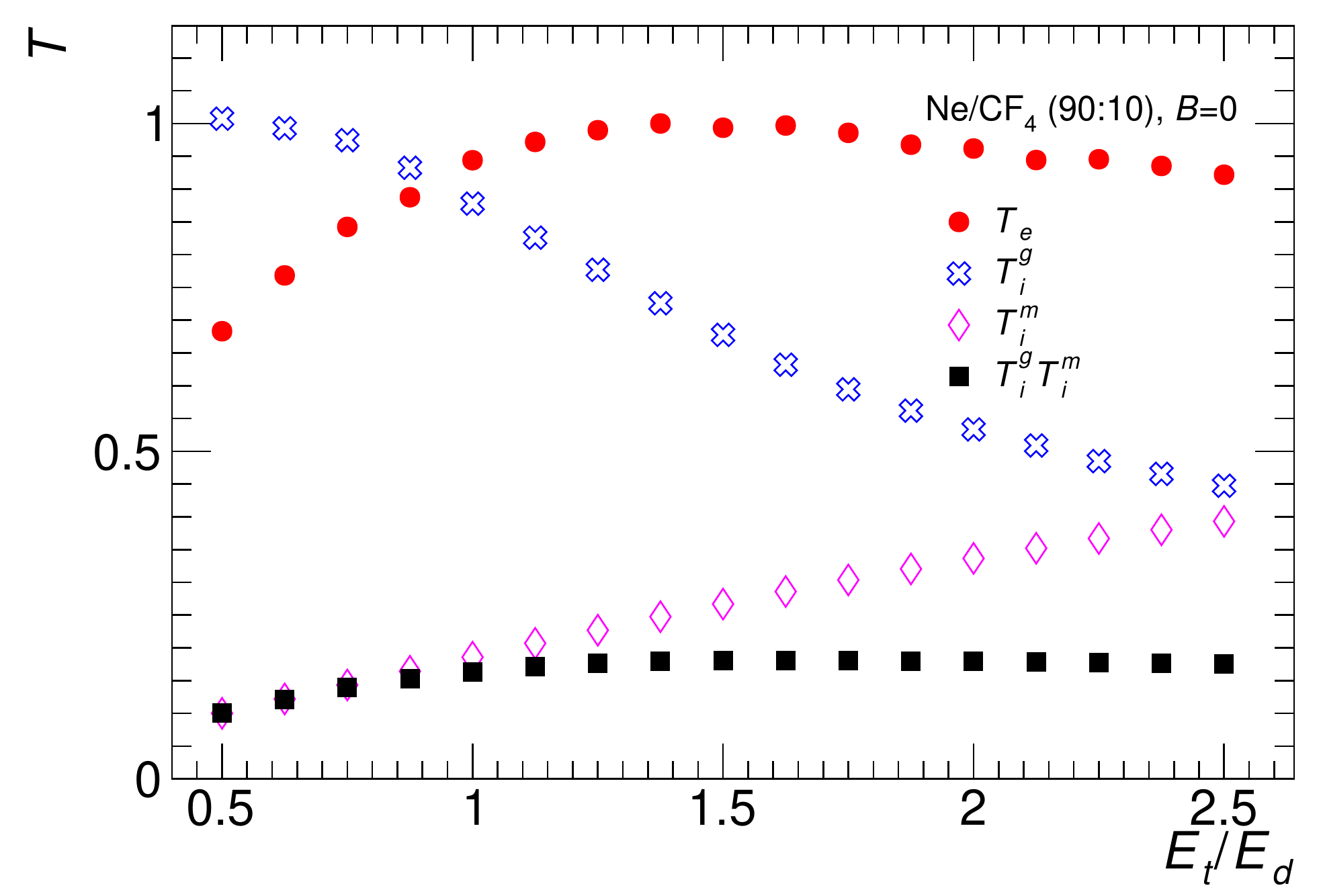}
   \caption{Transparencies as a function of $\Et/\Ed$ scan for $\Ed=320$~V/cm. \Te is the product of $\Teg\Tem$ normalized to unity at maximum.}
   \label{fig:edscan}
\end{figure}
The electron transparency curve \Te shown with circles is the product of $(\Teg\Tem)$. Since the electric field below \mesh is defined by the voltage on \anode wires and is much stronger than \Et, \Tem is high and therefore does not strongly depend on \Et, as long as \Et is low. A small decrease in \Te above $Et/Ed>1.5$ seen in the figure may indicate a departure from this regime. \Te rises with increasing \Et and reaches a maximum around $\Et=1.5\Ed$. \Tig, shown with crosses, steadily decreases with increasing \Et. This, however, is offset by an increase of \Tim shown in the plot with diamonds. Analogous effects would also be present in a real detector in which \Et coupled to the \TPC amplification plane below the \BPG, would extract ions into the drift volume of the detector. Those two effects nearly cancel each other above $\Et = \Ed$ as shown in Fig.~\ref{fig:edscan} with square symbols, which are the product of (\Tig\Tim).

As a result of this study the working setting is chosen $w=\Et/\Ed = 1.5$, $\Et = 480$~V/cm. Since the result measured in different gases are comparable, and following the decision to use the constant \Ed, the same \Et value is used in all measurements to facilitate comparisons between different gas mixtures.

\subsubsection{Charge measurement}
\label{sec:charge_measurement}
The \pad electrode is connected to the charge measurement line. To minimize the MCA dead time $<5\%$ the collimator is adjusted to provide the counting rate in the detector below $10^5$~s$^{-1}$. Three measurements are taken with fields set to: 
\begin{enumerate}
    \item $\Et = \Ed = 0$,
    \item $\Ed = 0$, $\Et = 480$~V/cm,
    \item $\Ed = 320$~V/cm, $\Et = 480$~V/cm.
\end{enumerate}
Data-taking time for each measurement is 5~min, the measurements are done to collect sufficient statistics. The results of this measurement are shown in the left panel of Fig~\ref{fig:charges} and are used for absolute normalization of the \Teg.

\subsubsection{\BPG voltage scan}

For this measurement \pad electrode is reconnected to the picoammeter, \cath, and \BPG are kept at the setting as for the last measurement in Sect.~\ref{sec:charge_measurement}, and the collimator is returned to its previous setting. The measurements are taken for \dV rising from 0 to 80~V in 2~V increments. The \pAs values are averaged over 60~s time after 10~s waiting time following each change in the voltage settings. Results of this measurement for $\Teg$ are shown in the right panel of Fig.~\ref{fig:charges} and for $\Tig$  in Fig.~\ref{fig:current_test}.

\section{Uncertainties of the measurements}
\label{sec:errors}

The nominal accuracy of the devices used in the measurement plays little role in the final results. These include precision of the power supplies, gas flow controllers, magnet, measuring devices, etc. 

The mechanical tolerance of the setup assembly is within hundreds of microns so that the fields discussed in the paper are known with a typical accuracy of 5\%. The non-uniformity of the gas gain, wire spacing, impact of the field distortion at the edges was studied by changing the illumination angle of the collimator radiation while keeping a similar counting rate. An approximate 2\% difference was found in the result which is assigned to the uncertainties.

Detector stability was estimated to contribute up to 5\% uncertainty which is the difference between two identical measurements performed with a month interval during which the detector was reassembled and the gas mixture was changed more than once.

Possible residual space-charge effects in the measurements performed with the currents $\ic<1$~nA are estimated as 10\% of the difference between the measurements done at $\ic=10$~nA and $\ic=1$~nA. They contribute up to 2\% at the highest \ic in the measurement of \Tig. 

The absolute normalization of the \Teg curve explained in Sect.~\ref{sec:charge_measurement} relies on the extrapolation of the curves shown in Fig.~\ref{fig:charges} to zero values. A 3\% uncertainty is added to the result based on the uncertainties in extrapolation.

As follows from the setup shown in Fig.~\ref{fig:setup}, applying \dV to \BPG shall not affect the \mesh transparencies as long as the average \BPG potential remains the same. However, a small change up to 7\% of the measured \Tim value given by Eq.~\ref{eqn:tim}, is observed in the experiment. This value is directly assigned to the \Tig as an uncertainty.

The contribution of different sources depends on \dV. At the highest transparency values, the uncertainties reach 7.5\%. At the transparency values close to zero the systematic uncertainty remains at the level of 0.5\% of the full scale and is approximately symmetric around zero, reflecting the fact that the results are obtained by subtracting measured currents as explained in Sect.~\ref{sec:mesurements}. The uncertainties of \Teg and \Tig are not the same, but are close in their values and are partially correlated with each other. To preserve the clarity of the plots, the dependence of systematic uncertainty on \dV is shown as a band around zero. It has to be taken as the uncertainty estimator for individual curves shown in figures.

\section{Results}
\label{sec:results}
The \BPG transparencies in Ne:CF$_{4}$ (90:10) gas mixture are shown in Fig.~\ref{fig:t_necf49010}. Results for other gas mixtures are given in the Appendix. 
\begin{figure}[htb]
   \centering
   \includegraphics*[width=0.49\textwidth]{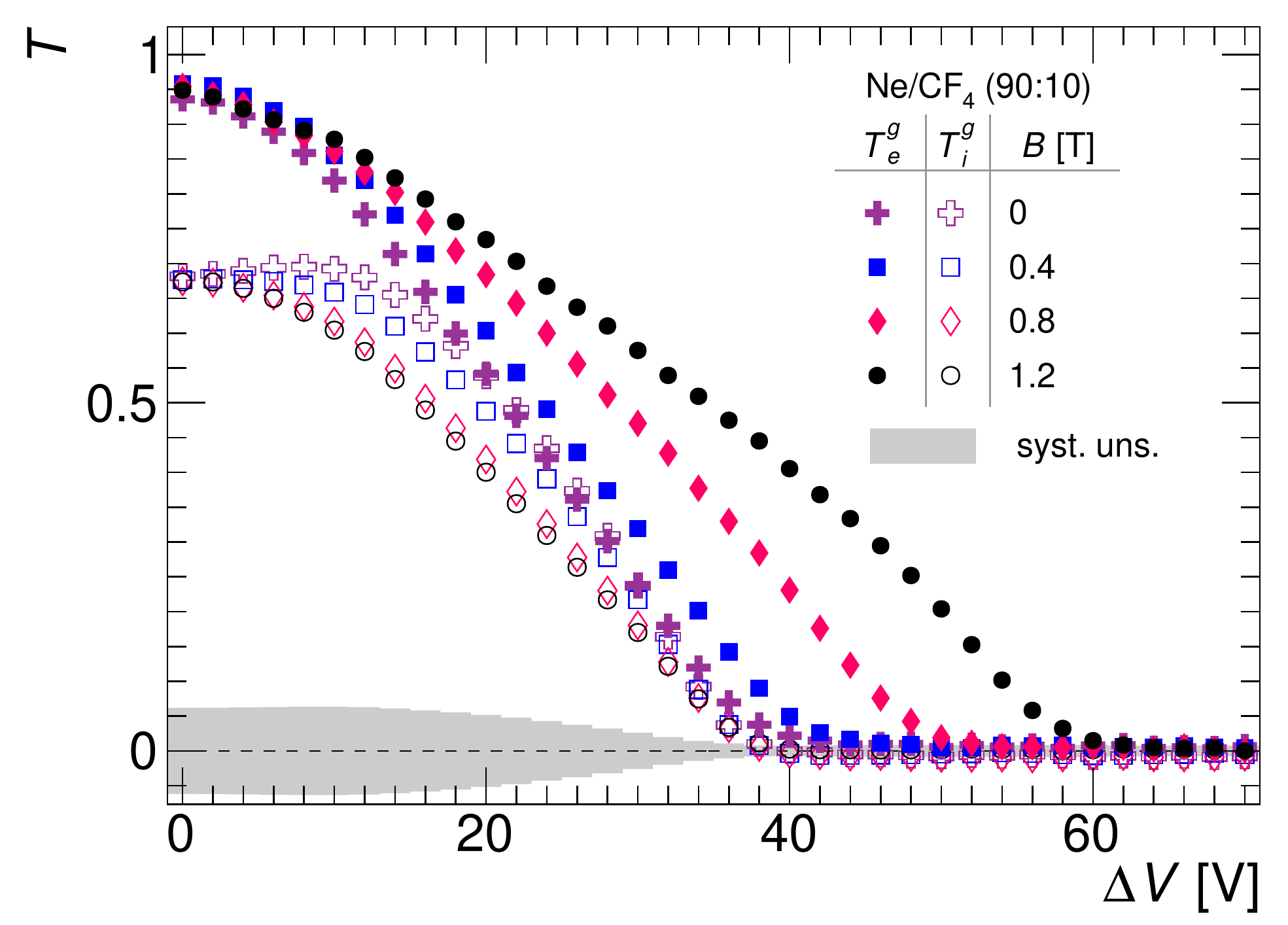}
   \caption{\BPG transparency as a function of \dV in Ne/CF$_{4}$ (90:10) gas mixture at different magnetic field setting. $\Ed=320$~V/cm, $\Et=480$~V/cm.  }
   \label{fig:t_necf49010}
\end{figure}
Measurements are done with the magnetic field switched off, and at the values 0.4~T, 0.8~T and 1.2~T. \Teg and \Tig are denoted by closed and open markers respectively. When the \BPG is at $\dV=0$ for all values of the magnetic field $\Teg\approx0.95$ and $\Tig\approx0.67$, which is defined by the choice of $\Ed/\Et$. With increasing \dV both transparencies decrease and reach zero. In the absence of the magnetic field, it occurs at $\dV\approx40$~V for ions and electrons. This voltage remains the same for ions also in the presence of the magnetic field although the shape of the \Tig changes around 10--30~V. This effect is not fully understood. In the presence of magnetic field \Teg behavior changes. It reaches zero at higher and higher voltages with increasing the magnetic field. At 1.2~T \Teg is still around 0.5 when \Tig is at zero. Thus, the \IBF can be fully shut at the expense of losing approximately half of the primary ionization. In the highest measured field setting the shape of the \Teg exhibits a kink at around $\dV=45$~V. Analogous behavior was also seen in~\cite{AMENDOLIA198547}. 

To quantitatively assess the insertion of the \BPG element into \TPC structure, one can introduce the figure of merit (\FoM) that is the ratio of the \IBF flowing into the \TPC with and without the \BPG. The \FoM depends on the ratio of the transfer and drift fields $w=\Et/\Ed=1.5$ discussed in Sect.~\ref{sec:et_measurement} and can be defined as:
\begin{eqnarray}
\FoM\left(w,\dV\right)=\frac{\Tim(w,0)}{\Tim(1,0)} \frac{\Tig(w,\dV)}{\Teg(w,\dV)}.
\label{eqn:fom}
\end{eqnarray}
The \FoM is the product of two terms. The first term results from the discussion of Fig.~\ref{fig:edscan}, that higher \Et extracts more ions from the amplification plane of the \TPC. Ion current extracted from that plane is characterized in Eq.~\ref{eqn:fom} by the ion transparency of the \mesh. Thus, the first term is the ratio of \Tim at the working setting of $w=1.5$ to that at $w=1$. The latter corresponds to the setup without the \BPG in which the \Ed is coupled directly to the \mesh.

The second term is the ion transparency \Tig, divided by the electron transparency \Teg. Denominator reflects the fact that the loss of primary ionization in \BPG must be compensated by raising the gain in the \TPC readout plane, which in turn would generate more ions. The \FoM defined by Eq.~\ref{eqn:fom} is greater than \Tig for any \dV. A \TPC with the \BPG has better performance when the \FoM takes smaller values.

Figure~\ref{fig:fom_vs_teg} showing \FoM as a function of \Teg demonstrates how much the  \IBF in a \TPC can be suppressed by introducing the \BPG element into its structure.
\begin{figure}[htb]
\centering
\includegraphics*[width=0.49\textwidth]{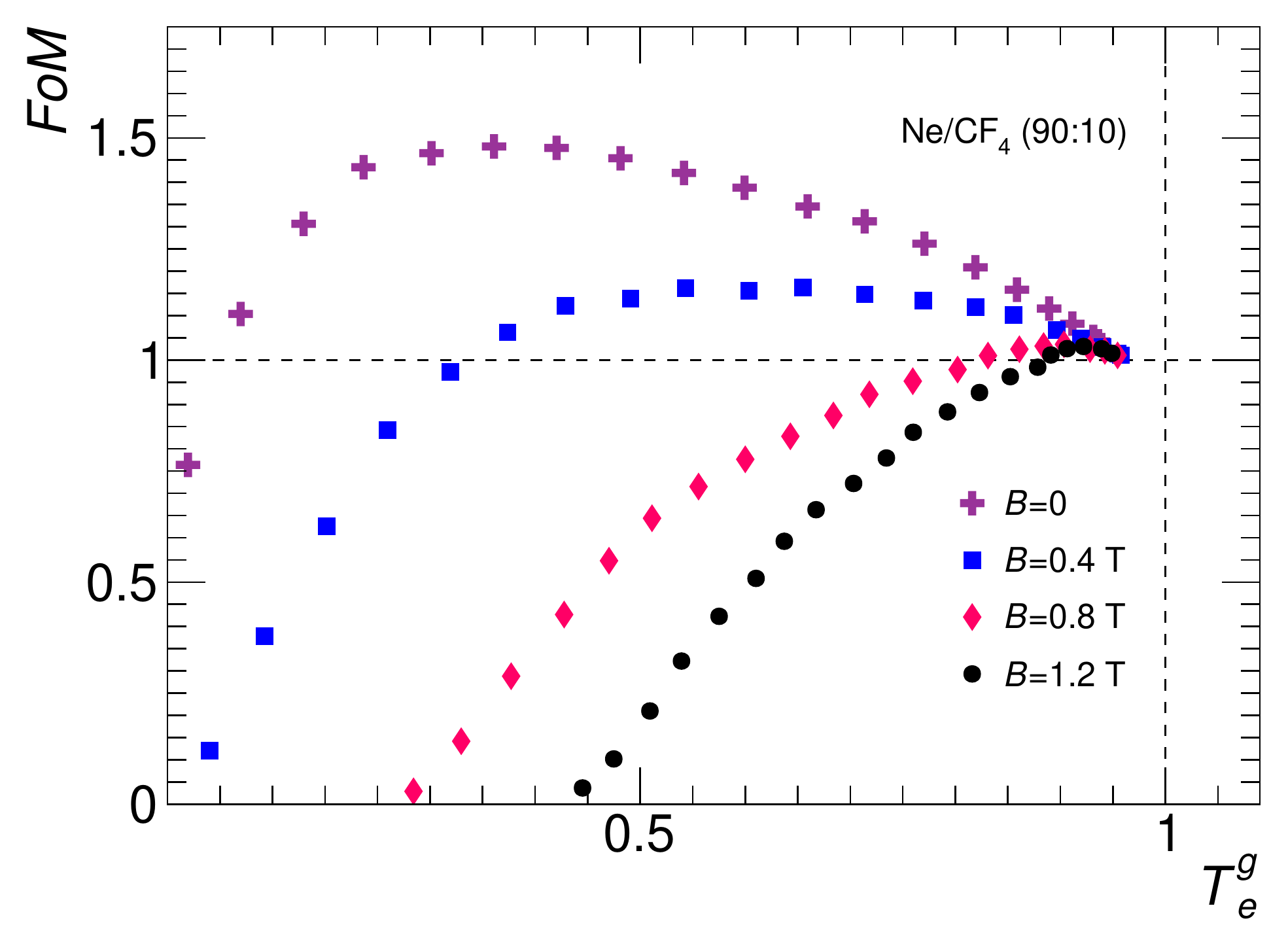}
\caption{\FoM vs. \Teg for different magnetic field settings.}
\label{fig:fom_vs_teg}
\end{figure}
As follows from Eq.~\ref{eqn:fom}, point (1,1) indicated in the figure by a crossing of dashed lines, corresponds to the case when the \BPG is absent in a \TPC. All graphs in the figure start in the vicinity of point (1,1) at $\dV=0$. This shows that the \BPG at a constant voltage makes little change to the \TPC performance in any magnetic field. Rising \dV on the \BPG wires in a low magnetic field leads to the loss of primary ionization and increase of the \IBF, which is seen in the figure from the curves remaining above unity even at low \Teg. The situation rapidly improves with the increase of the magnetic field, in which the \BPG effectively suppresses the \IBF while keeping most of the primary ionization.

Suppression of the \IBF by the \BPG leads to loss of primary electron ionization and thus deteriorates the \TPC $dE/dx$ resolution~\cite{Abelevetal:2014cna}. In the case of the \BPG, this effect can be estimated in its leading order as a loss of primary electron statistics. Assuming that the relative loss of the $dE/dx$ resolution is reciprocal to $\sqrt{\Teg}$ one can plot it vs. the \FoM as shown in  Fig.~\ref{fig:dedx_vs_fom}. 
\begin{figure}[htb]
   \centering
   \includegraphics*[width=0.49\textwidth]{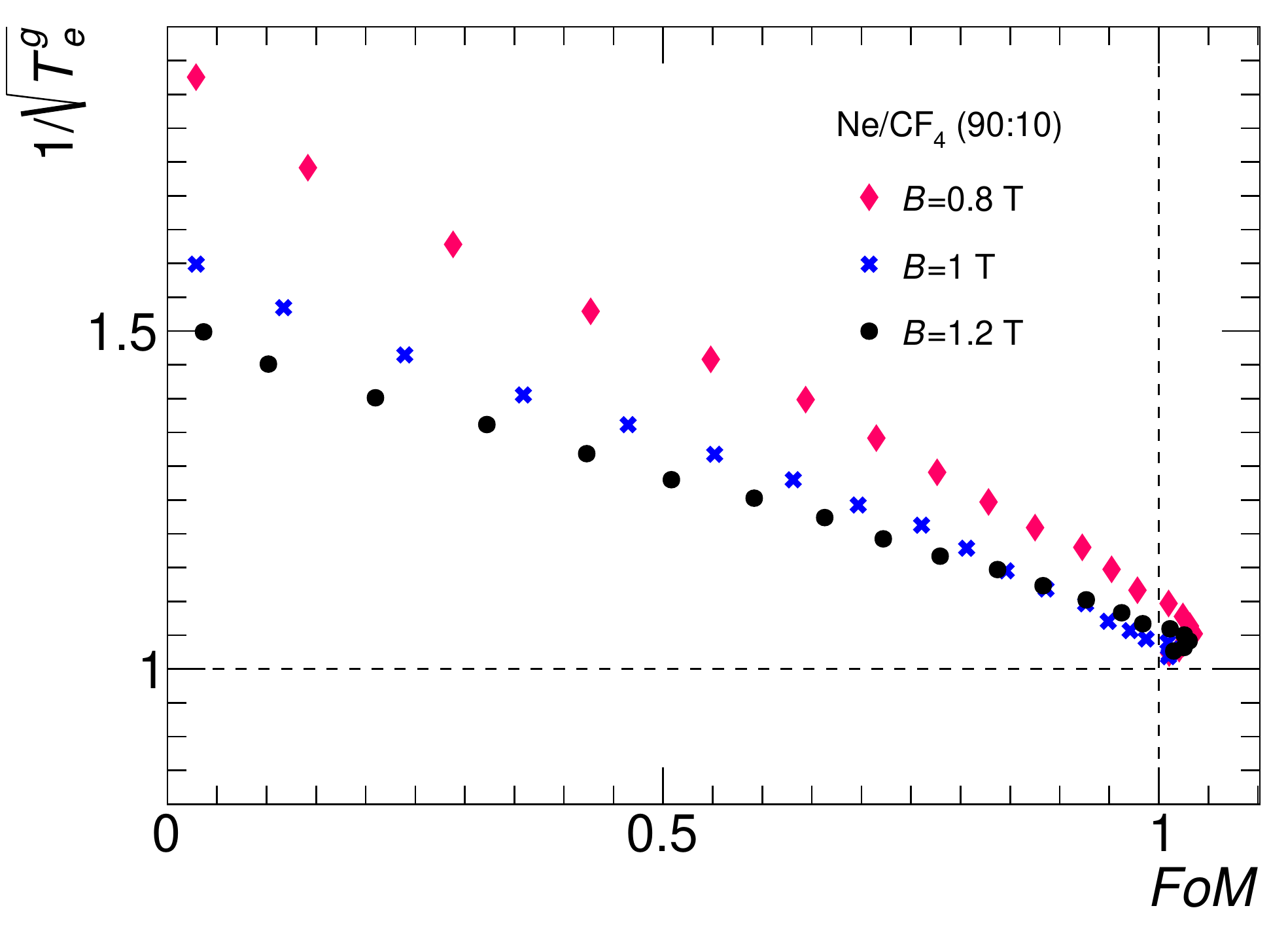}
   \caption{Dependence of the TPC $dE/dx$ resolution proportional to $1/\sqrt{\Teg}$ vs. \FoM at high magnetic fields. Values compare to case of no \BPG in the detector, corresponding to point (1,1).}
   \label{fig:dedx_vs_fom}
\end{figure}
The curve measured in the $B=1.2$~T shows that $\IBF$ suppression by a factor of 5 is achievable at the expense of $40\%$ deterioration in the $dE/dx$ resolution, and it can be fully suppressed at a cost of 55\% of the resolution loss. If such \BPG is installed in a \TPC with the $dE/dx$ resolution of 10\% and \IBF of 2\%, the resulting performance would be given by the product of these numbers: the $\TPC\&\BPG$ configuration would have $dE/dx$ resolution of 14\% at $\IBF=0.4$\% and no \IBF at the $dE/dx$ resolution of 15.5\%. The trend of the curve measured at $B=1.2$~T improves in a higher magnetic field.

Garfield++ toolkit~\cite{Garfield} is used to simulate the \BPG performance. Gas properties are simulated for Ne/CF$_{4}$ (90:10) and Ar/CH$_{4}$ (90:10) to reproduce the experimental conditions. Detector electrodes are modeled using {\it ComponentAnalyticField} class. Electrons are injected 2~mm above the \BPG and ions originate from the volume of 60~$\mu$m in diameter around \anode wires. The simulations use {\it DriftLineRKF} class to calculate the drift lines of the particles. Figure~\ref{fig:mc_to_data} compares simulation to the measured 
\begin{figure*}[htb]
   \centering
   \includegraphics*[width=0.49\textwidth]{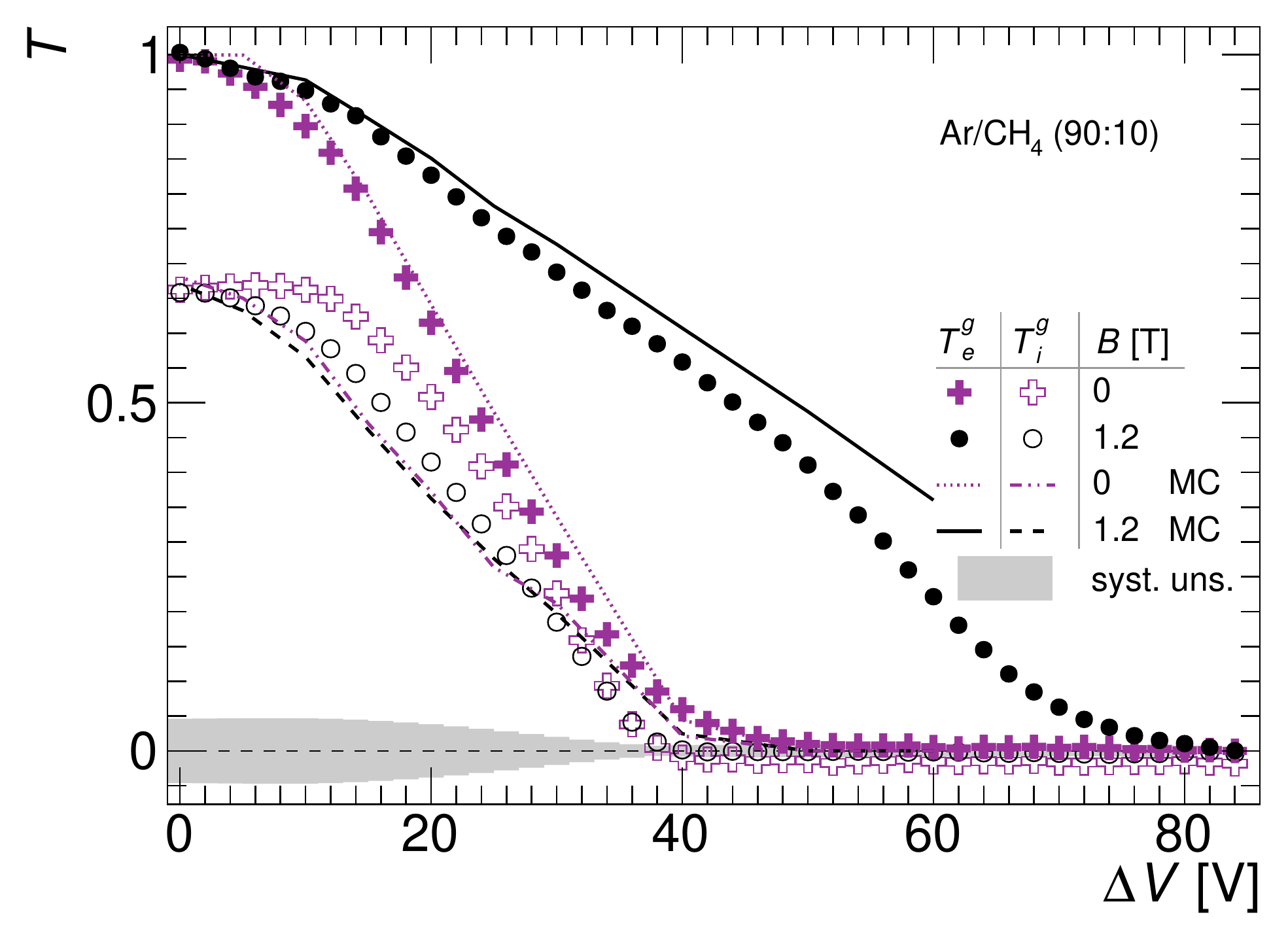}
   \includegraphics*[width=0.49\textwidth]{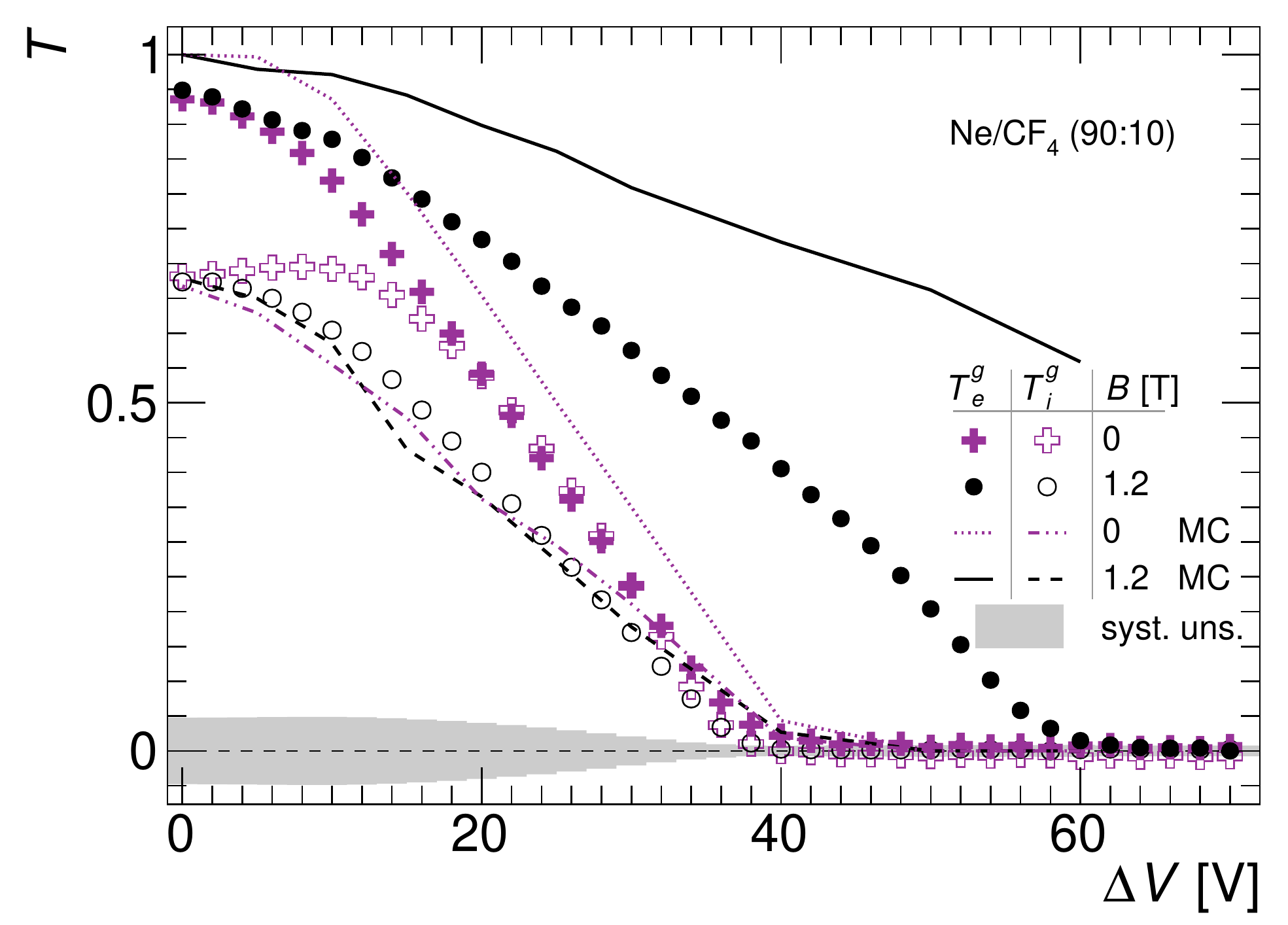}   
   \caption{Transparencies as a function of \dV compared to the Garfield++ simulation for Ar/CH$_{4}$ (90:10) (left) and Ne/CF$_{4}$ (90:10) (right) gas mixtures.}
   \label{fig:mc_to_data}
\end{figure*}
results for Ar/CH$_{4}$ gas mixture shown in the left panel and Ne/CF$_{4}$ shown in the right panel. Comparisons are done for settings without a magnetic field and for $B=1.2$~T. In the absence of the magnetic field, the simulation reproduces the electron and ion data within approximately 10\%, comparable to the data measurement accuracy. In the presence of a magnetic field, the simulation curves for ions remain the same, whereas the data shows different shapes. Nevertheless, the point where the \Ti reaches zero is well reproduced by the simulation. For Ar/CH$_{4}$ gas mixture the simulation curve for \Te agrees with the data reasonably well but for Ne/CF$_{4}$ it shows the significant deviation. 
\section{Discussion and conclusions}
\label{sec:conclusions}
The experimental setup built at the Weizmann Institute of Science is used to measure electron and ion transparencies of a bipolar wire grid operated in a magnetic field in passive mode. Studies are made in Ne-based and Ar-based gas mixtures using CH$_4$, CF$_4$, and CO$_2$ as quenchers. The results for Ar/CH$_4$ (90:10) are qualitatively consistent with the measurements published in Ref.~\cite{AMENDOLIA198547}. 

The performance of the bi-polar grid is evaluated in terms of transparencies to electron and ion currents traversing it from above and from below respectively. Since the transparencies of the grid strongly depend on the electric fields coupled to it (Fig.~\ref{fig:fom_vs_teg_ArCH4_90_10}~vs~\ref{fig:fom_vs_teg_ArCH4_90_10_Et0p5}), most measurements are performed in configuration with electric fields 320~V/cm and 480~V/cm above and below the grid respectively. This configuration is chosen to facilitate comparisons between different gas mixtures. As a result, several common features can be seen in all measurements (Figs.~\ref{fig:t_necf49010},~\ref{fig:fom_vs_teg},~\ref{fig:fom_vs_teg_NeCF4_50_50}--\ref{fig:fom_vs_teg_ArCO2_90_10}). Without voltage bias on the grid wires, grid transparency to ions is about 70\% and grid transparency to electrons is above 90\% in all gas mixtures, except in Ar/CO$_2$ (Fig.~\ref{fig:fom_vs_teg_ArCO2_90_10}), where it is close to the ion transparency. Transparency values for electrons and ions measured at zero bias in the main field configuration do not depend on the strength of the magnetic field. Increasing voltage bias on the wires to $\pm$40~V in all gases brings the ion transparency to zero even in the strongest measured magnetic field of 1.2~T. At 1~mm pitch between the grid wires, this bias corresponds to an electric field of approximately 800~V/cm, twice the average of the coupled fields. An empirical estimate that the field inside the grid required to zero out the ion current through it shall be twice the field coupled to the grid also holds for other field configurations measured in this study (Figs.~\ref{fig:fom_vs_teg_ArCH4_90_10_Et0p5}, \ref{fig:fom_vs_teg_ArCH4_90_10_Ed140}). The shape of the curve for ion transparency shows weak dependence on the magnetic field, although the nature of the elevation that develops in the 10--30~V region in stronger magnetic fields is not clear. Garfield-based simulations well reproduce the grid transparency to ions but show no shape dependence on the magnetic field.

Grid transparency to electrons is sensitive to the magnetic field and the required voltage bias to zero out the electron current increases in higher magnetic field by nearly a factor of 2 compared to ions (Figs.~\ref{fig:fom_vs_teg_NeCH4_90_10},~\ref{fig:fom_vs_teg_ArCH4_90_10}). At $\pm$40~V required to stop the ion flow, in 1.2~T field the grid retains 45--60\% transparency to electrons. The simulations reproduce the behavior of grid transparency to electrons in the absence of a magnetic field, but in some gases, the simulations show significant deviations from the measured curves when the magnetic field is present (Fig.~\ref{fig:mc_to_data}).

To quantitatively evaluate the impact of the grid element in the structure of the \TPC a figure of merit is introduced as explained in Sect~\ref{sec:results}. Its smaller values correspond to better performance of the \TPC with the grid in suppressing positive ion backflow. A grid without voltage bias on its wires makes almost no impact on the \TPC performance in any magnetic field, except in Ar/CO$_2$ gas mixture. Although this may be seen as a trivial statement, the measurements show if a grid plane is built in a \TPC for the purpose to decouple drift and amplification regions, it makes almost no impact on the \TPC performance (Figs.~\ref{fig:fom_vs_teg},~\ref{fig:fom_vs_teg_NeCF4_50_50}--\ref{fig:fom_vs_teg_ArCO2_90_10}).

With the voltage bias on the wires, the grid performance strongly depends on the magnetic field. The effect of the grid in gas mixtures with small Lorentz angle using CO$_2$ as a quencher is insignificant, but it drastically improves in gases with larger Lorentz angles, such as the mixtures with CF$_4$ and CH$_4$. Between these two gases, CH$_4$ shows slightly better results (Fig.~\ref{fig:fom_vs_teg_NeCH4_90_10}~vs~\ref{fig:fom_vs_teg}, and Fig.~\ref{fig:fom_vs_teg_NeCH4_50_50}~vs~\ref{fig:fom_vs_teg_NeCF4_50_50}). The results somewhat improve in the mixtures with a lower concentration of the quenching gas (Fig.~\ref{fig:fom_vs_teg_NeCH4_90_10}~vs~\ref{fig:fom_vs_teg_NeCH4_50_50}, Fig.~\ref{fig:fom_vs_teg}~vs~\ref{fig:fom_vs_teg_NeCF4_50_50}, and Fig.~\ref{fig:fom_vs_teg_ArCH4_90_10}~vs~\ref{fig:fom_vs_teg_ArCH4_50_50}). For the same quencher, Ar-based mixtures show better results compared to Ne-based mixtures (Fig.~\ref{fig:fom_vs_teg_ArCH4_90_10}~vs~\ref{fig:fom_vs_teg_NeCH4_90_10}, and Fig.~\ref{fig:fom_vs_teg_ArCH4_50_50}~vs~\ref{fig:fom_vs_teg_NeCH4_50_50}). 

Although the results measured in this study are qualitatively consistent with the expectations coming from the theory of electrons and ions drift in gases~\cite{Hilke_2010,Sauli:1992bu}, the quantitative comparison shows significant deviations from the measured data, especially for electrons. Simulations based on Garfield++ toolkit are not sufficiently accurate in describing measurements in some gases (Fig.~\ref{fig:mc_to_data}).

To get more insight into this problem, measurement were also done in other field configuration (Figs.~\ref{fig:fom_vs_teg_ArCH4_90_10_Et0p5},~\ref{fig:fom_vs_teg_ArCH4_90_10_Ed140}). Field change results in different behaviour seen in curves, which in some cases are consistent with expectations determined by the field changes. A surprising result is that although in Ar/CH$_4$ the Lorentz angle in lower electric field is expected to almost double~\cite{BITTL1997249} compared to the main setting, the measurements show that it results into small reduction of transparency to electrons (Figs.~\ref{fig:fom_vs_teg_ArCH4_90_10}~vs~\ref{fig:fom_vs_teg_ArCH4_90_10_Ed140}).

Although some of the measured effects are not reproduced by simulation, results reported in the paper demonstrate that a passive bipolar grid operated in a magnetic field above 1~T can be used as an effective instrument to suppress the ion backflow in \TPCs.
\section{Acknowledgments}

The authors would like to express gratitude to Prof. Amos Breskin and Dr. Arindam Roy of the Weizmann Institute for fruitful discussions and help. The authors are thankful to the High-Energy group at the Weizmann Institute working on the ATLAS experiment for their help in making parts of the setup. The authors would like to thank the group of Prof. Mirko Planini\'{c} from the University of Zagreb for help with setting up floating picoammerts and the reliable performance of their devices during the measurements. Special thanks are to the Physics Core Facility group of the Weizmann Institute for their constant support during the experiment.
\appendix{}
\label{sec:svalka}

Measurements of \Teg and \Tig vs \dV and \FoM vs \Teg for the \BPG in different gases at zero magnetic fields and 0.4~T, 0.8~T, and 1.2~T. Measurements for Ne/CF$_{4}$ (90:10) gas mixture are given in Figs.~\ref{fig:t_necf49010} and \ref{fig:fom_vs_teg}. To preserve the clarity of the plots, the dependence of systematic uncertainty on \dV is shown as a band around zero. It has to be taken as the uncertainty estimator for individual curves shown in figures. The systematic uncertainties are described in Sect.~\ref{sec:errors}.

\begin{figure*}[htb] 
\centering
\includegraphics*[width=0.49\textwidth]{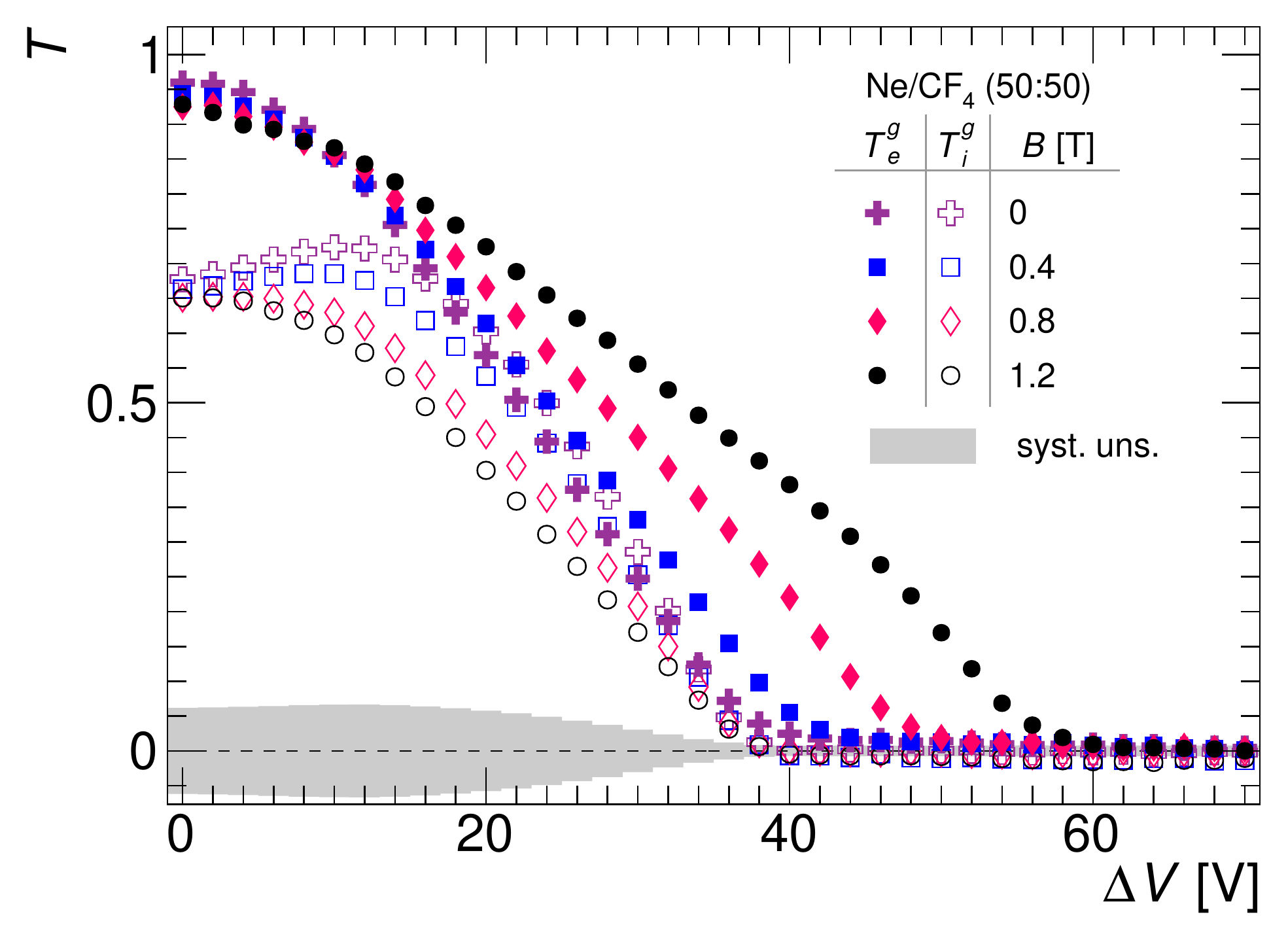}
\includegraphics*[width=0.49\textwidth]{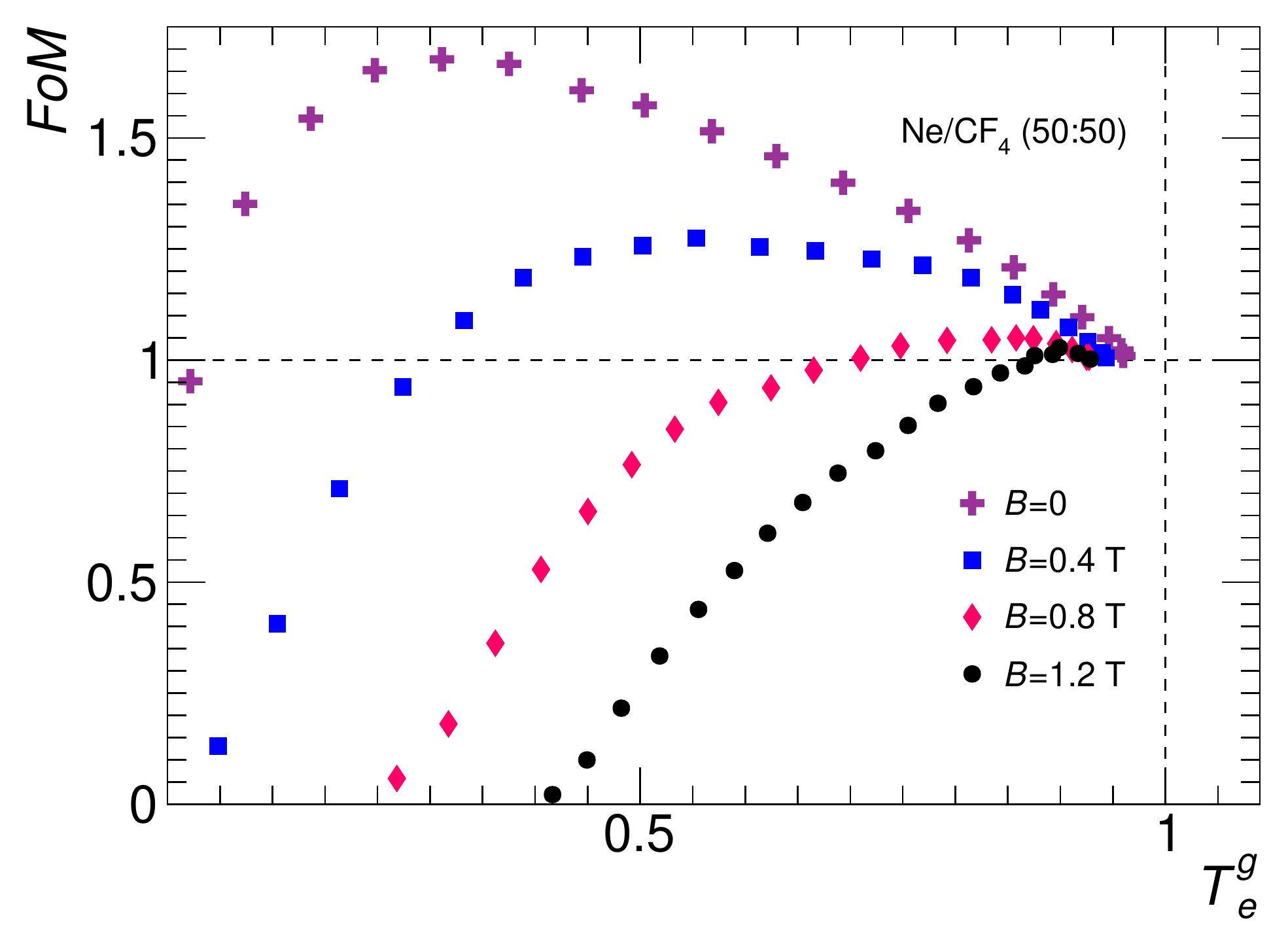}
\caption{\BPG performance in Ne/CF$_{4}$ (50:50) gas mixture at $\Ed$=320~V/cm, $\Et$=480~V/cm. Left: Transparency as a function of \dV. Right: \FoM vs. \Teg.}
\label{fig:fom_vs_teg_NeCF4_50_50}
\end{figure*}

\begin{figure*}[htb] 
\centering
\includegraphics*[width=0.49\textwidth]{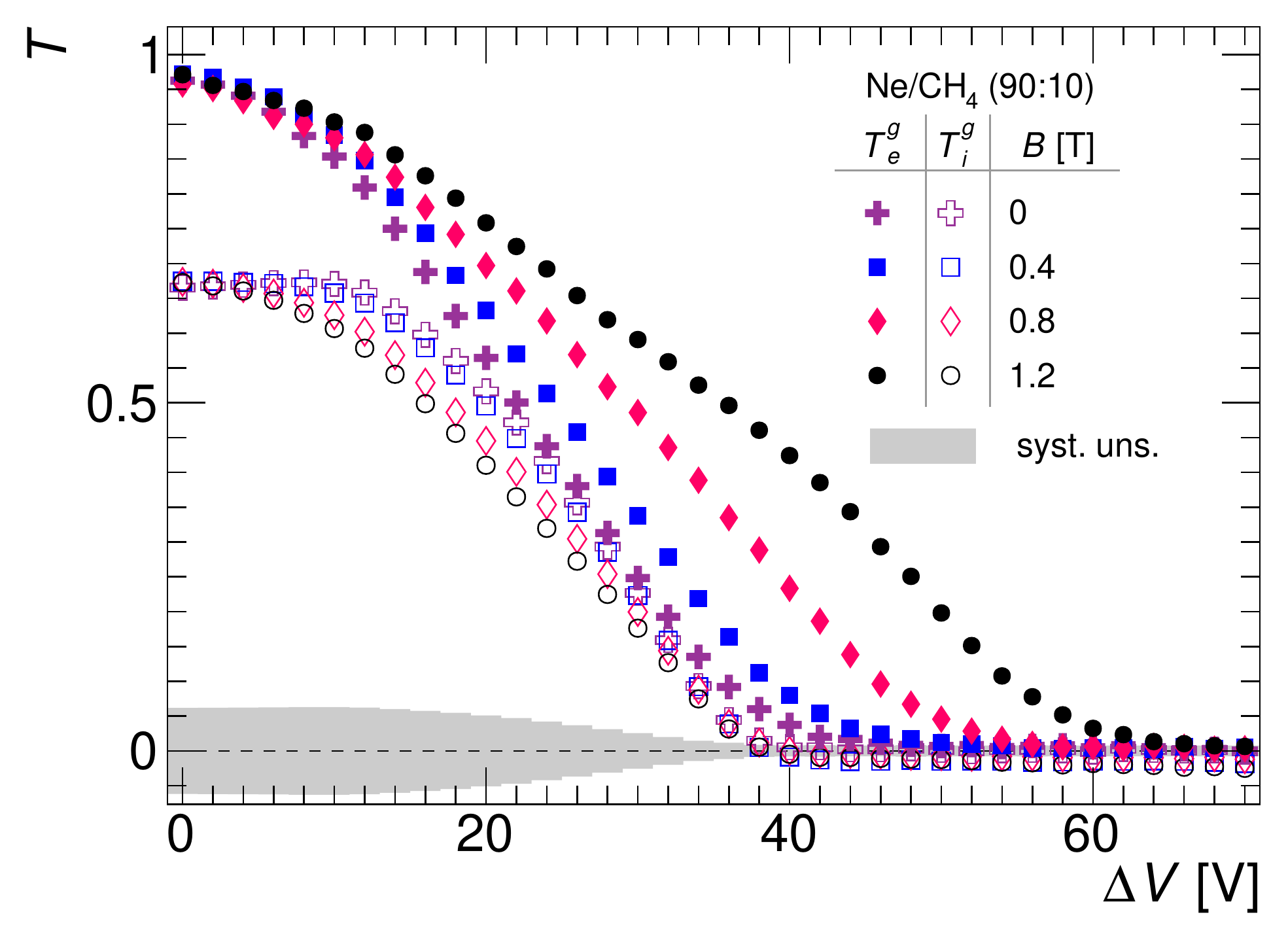}
\includegraphics*[width=0.49\textwidth]{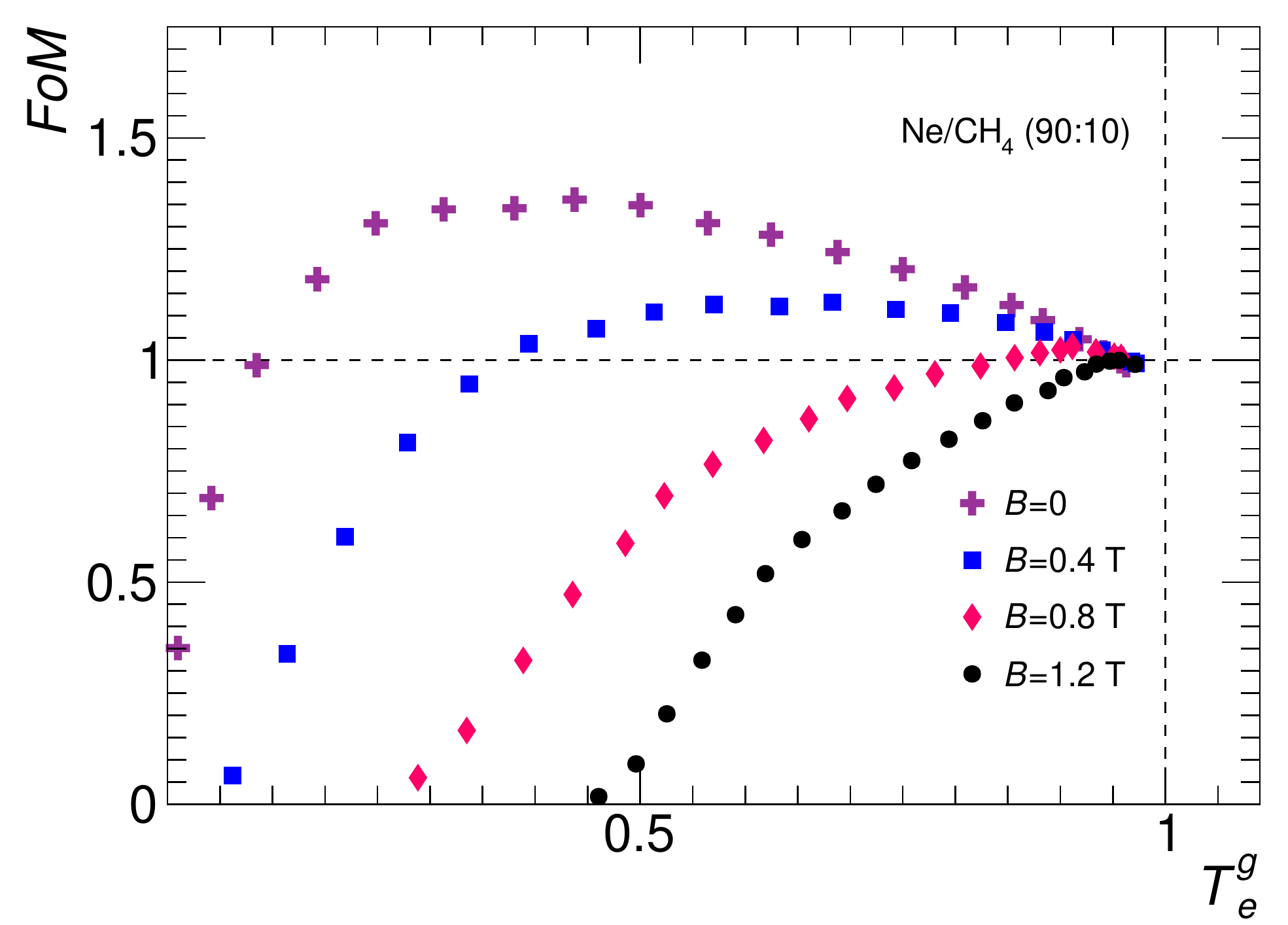}
\caption{\BPG performance in Ne/CH$_{4}$ (90:10) gas mixture at $\Ed$=320~V/cm, $\Et$=480~V/cm. Left: Transparency as a function of \dV. Right: \FoM vs. \Teg.}
\label{fig:fom_vs_teg_NeCH4_90_10}
\end{figure*}

\begin{figure*}[htb] 
\centering
\includegraphics*[width=0.49\textwidth]{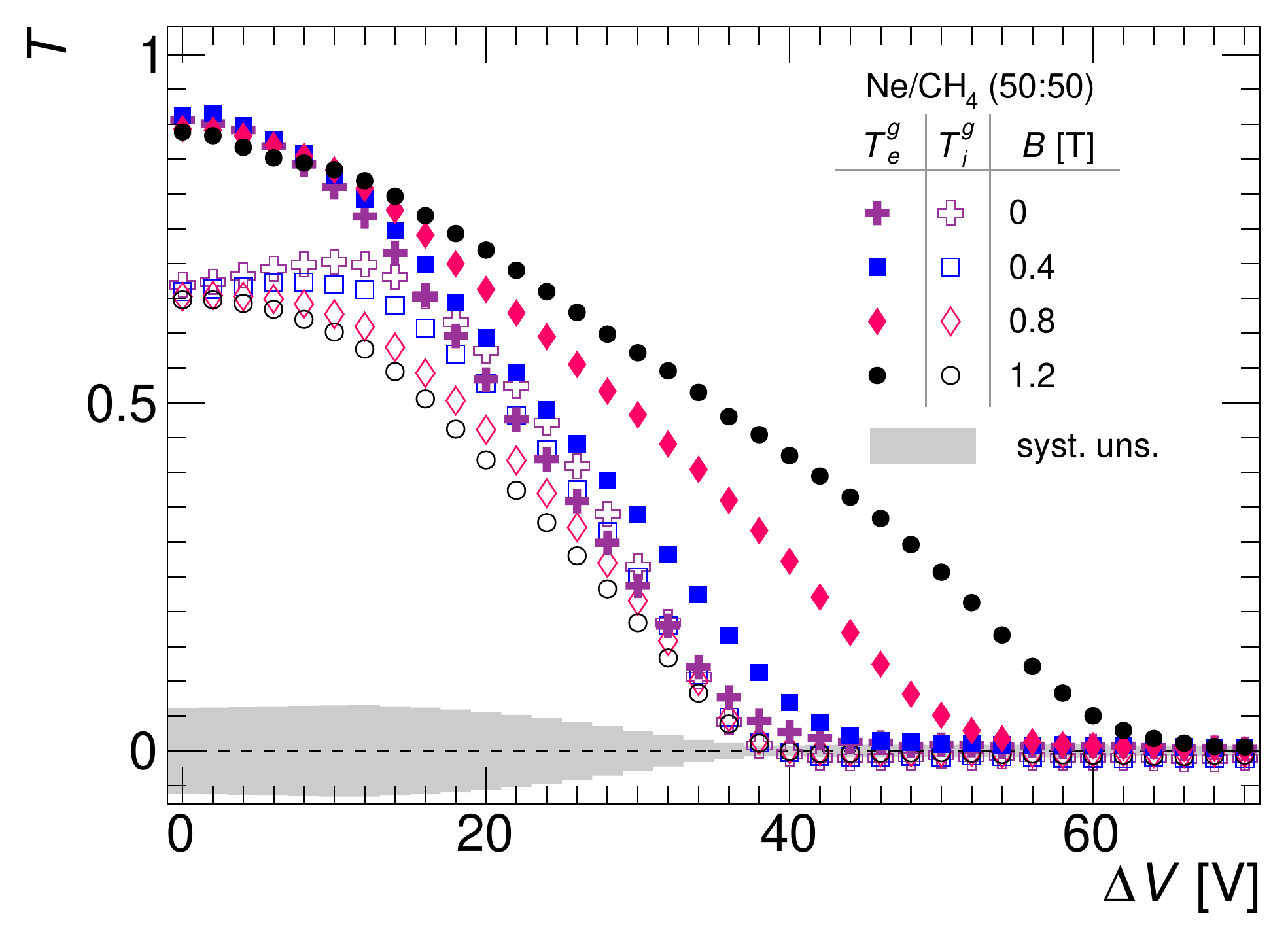}
\includegraphics*[width=0.49\textwidth]{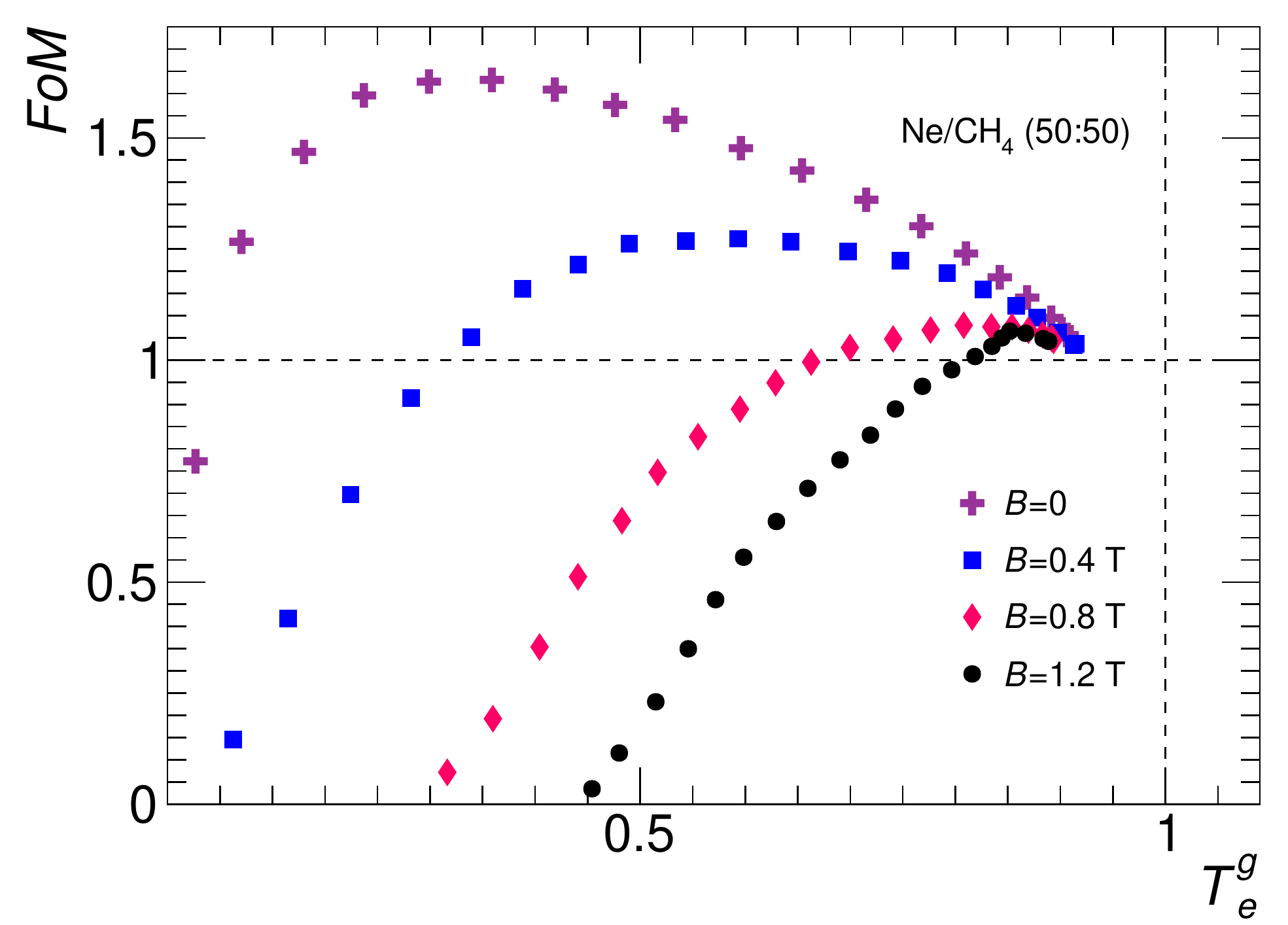}
\caption{\BPG performance in Ne/CH$_{4}$ (50:50) gas mixture at $\Ed$=320~V/cm, $\Et$=480~V/cm. Left: Transparency as a function of \dV. Right: \FoM vs. \Teg.}
\label{fig:fom_vs_teg_NeCH4_50_50}
\end{figure*}

\begin{figure*}[htb] 
\centering
\includegraphics*[width=0.49\textwidth]{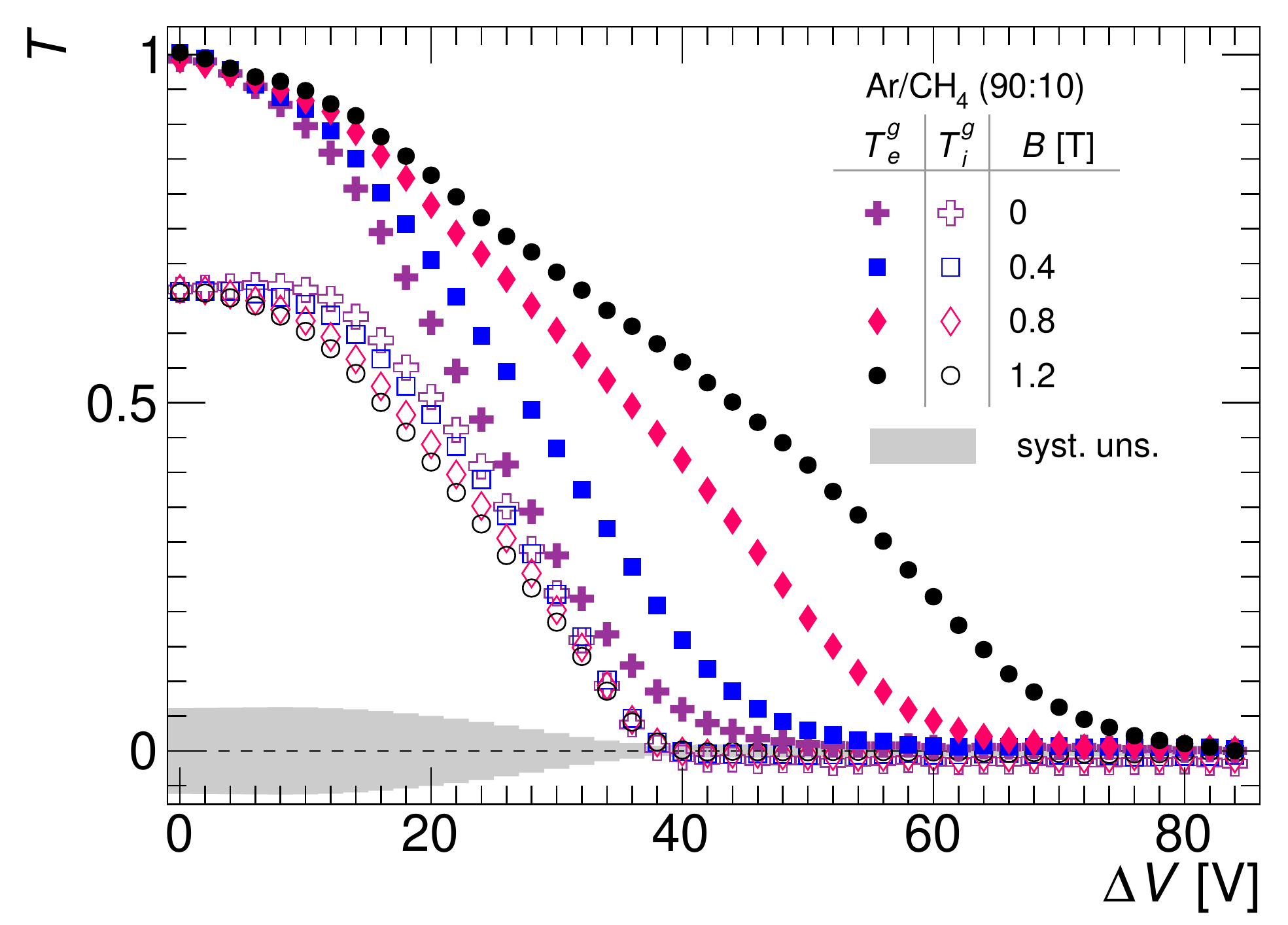}
\includegraphics*[width=0.49\textwidth]{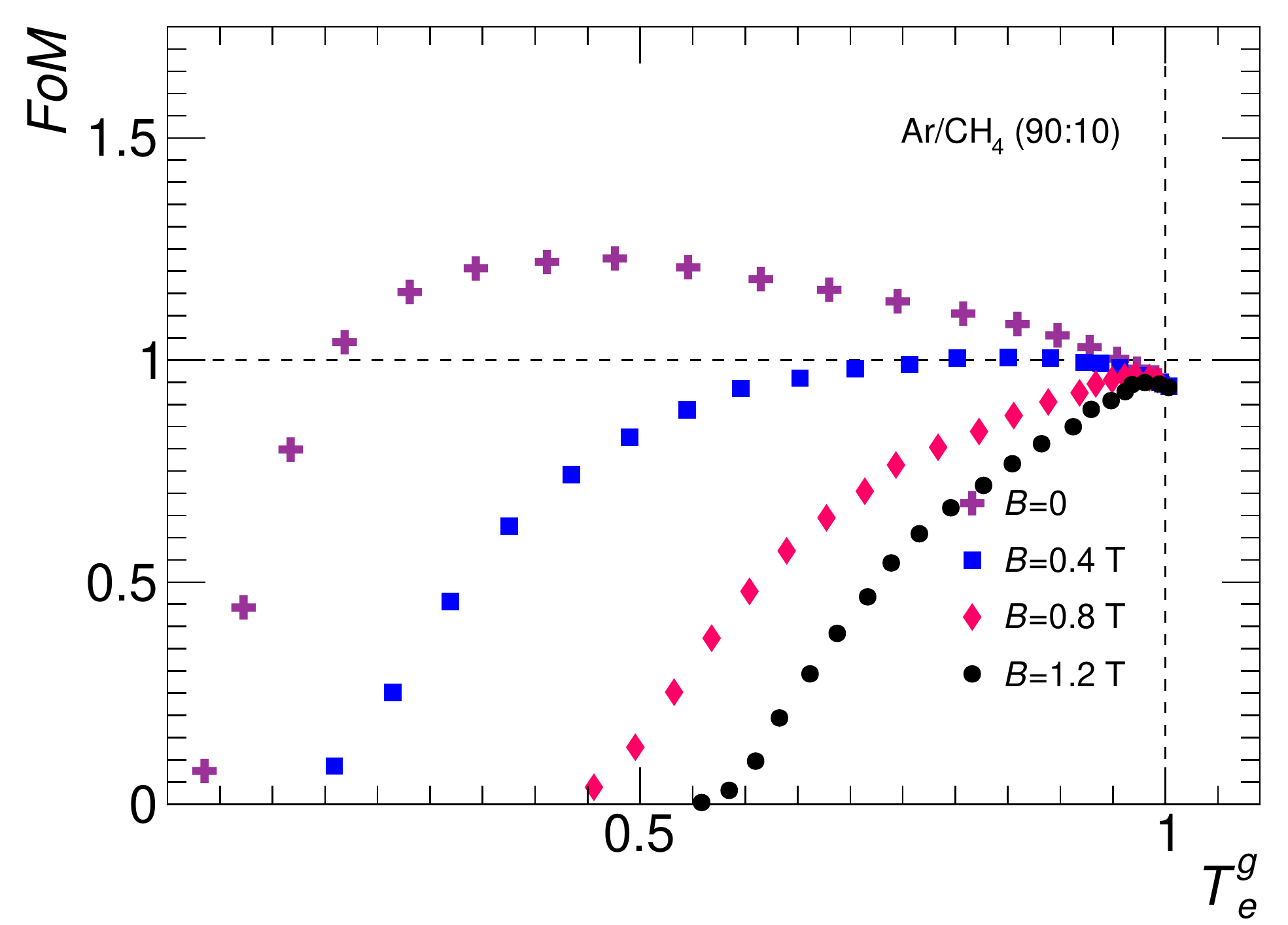}
\caption{\BPG performance in Ar/CH$_{4}$ (90:10) gas mixture at $\Ed$=320~V/cm, $\Et$=480~V/cm. Left: Transparency as a function of \dV. Right: \FoM vs. \Teg.}
\label{fig:fom_vs_teg_ArCH4_90_10}
\end{figure*}

\begin{figure*}[htb] 
\centering
\includegraphics*[width=0.49\textwidth]{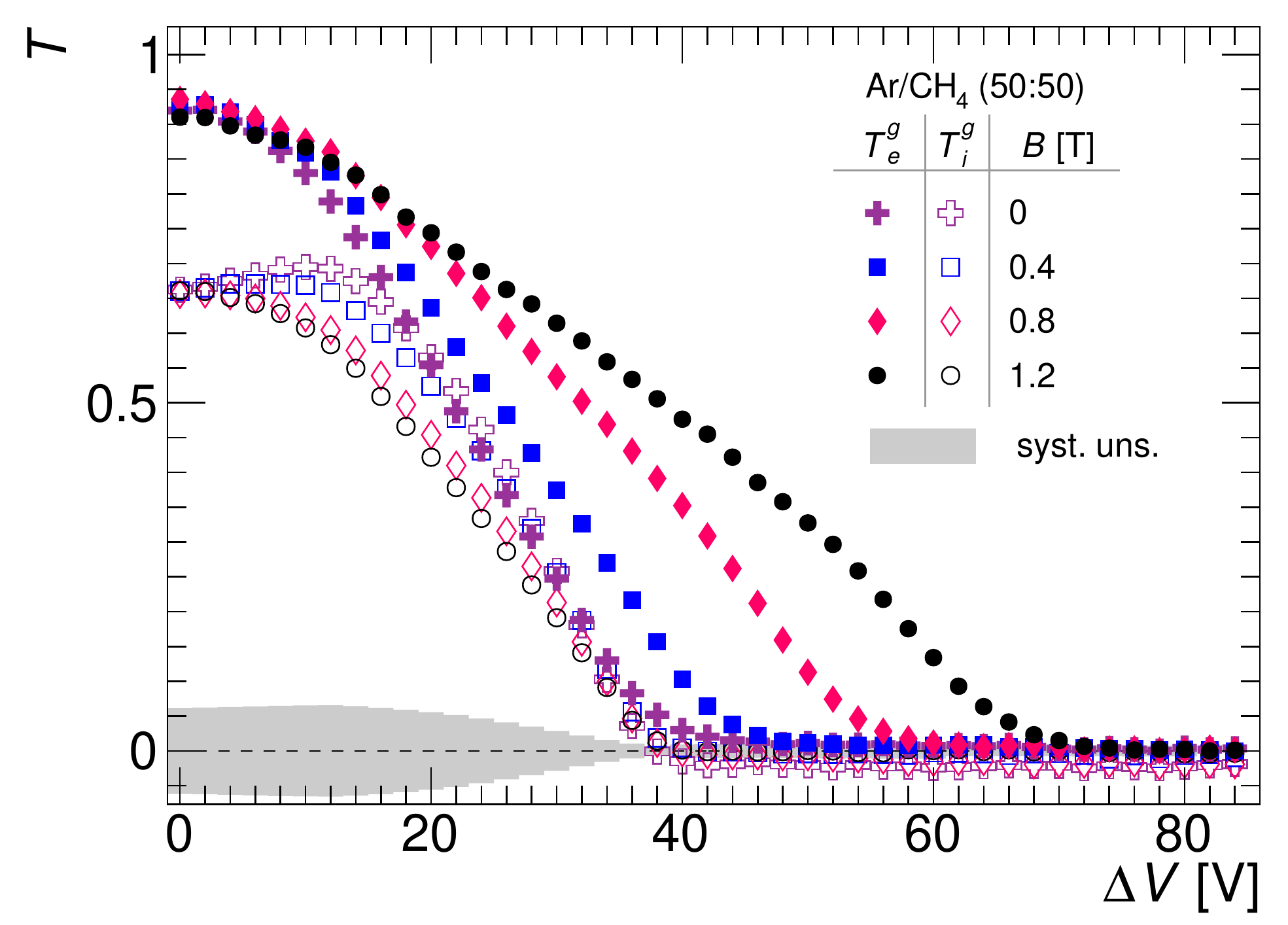}
\includegraphics*[width=0.49\textwidth]{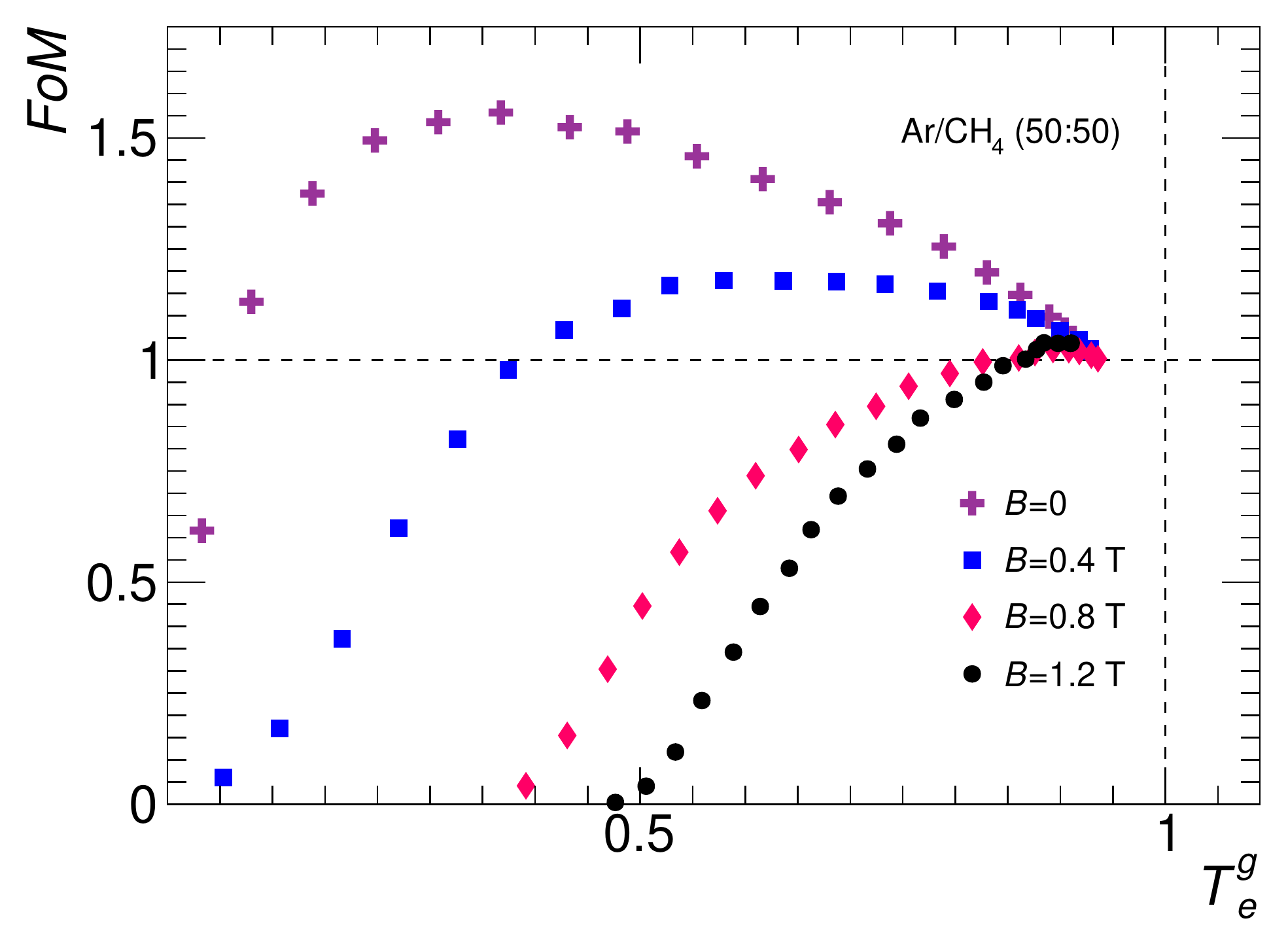}
\caption{\BPG performance in Ar/CH$_{4}$ (50:50) gas mixture at $\Ed$=320~V/cm, $\Et$=480~V/cm. Left: Transparency as a function of \dV. Right: \FoM vs. \Teg.}
\label{fig:fom_vs_teg_ArCH4_50_50}
\end{figure*}

\begin{figure*}[htb] 
\centering
\includegraphics*[width=0.49\textwidth]{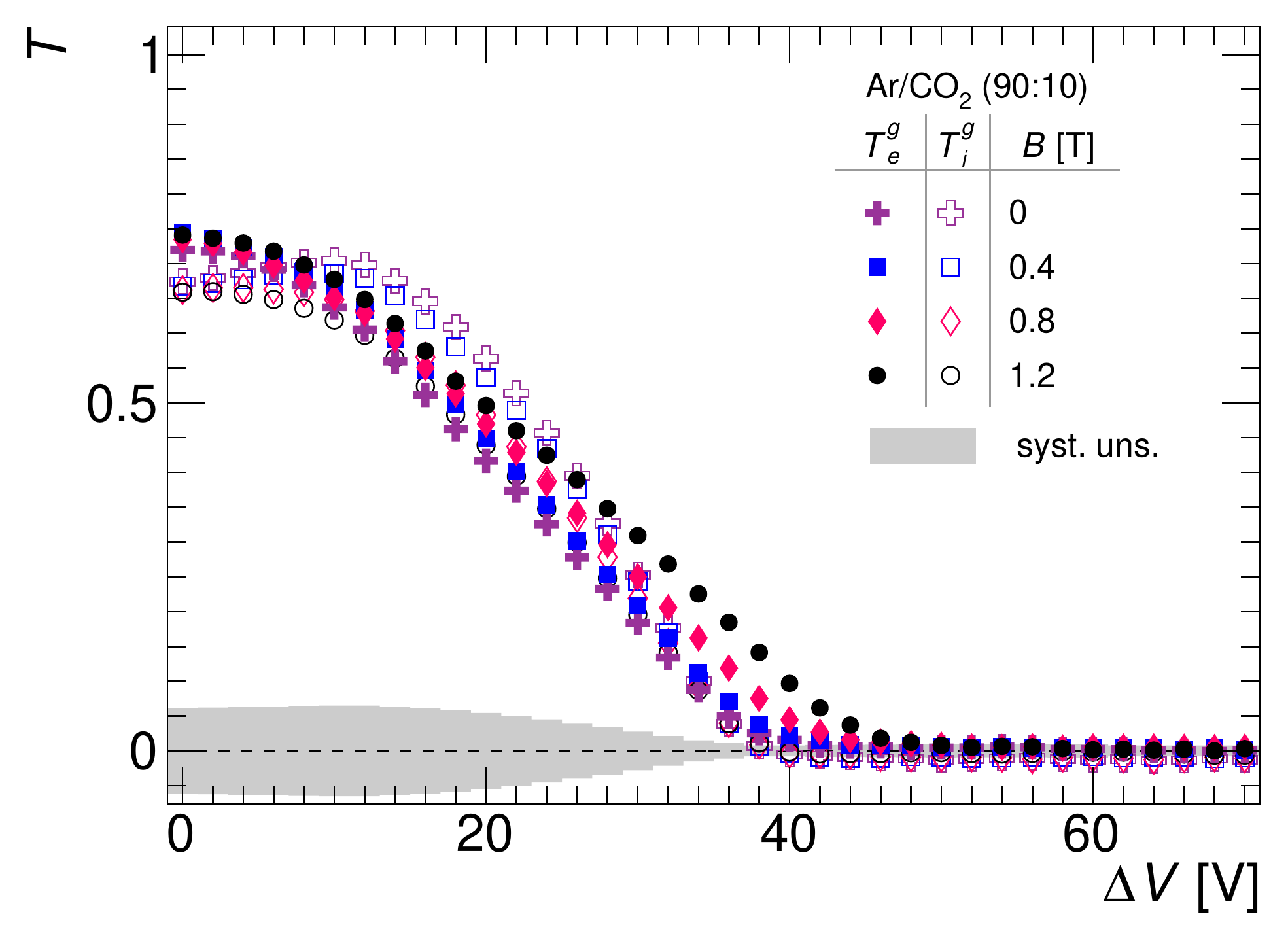}
\includegraphics*[width=0.49\textwidth]{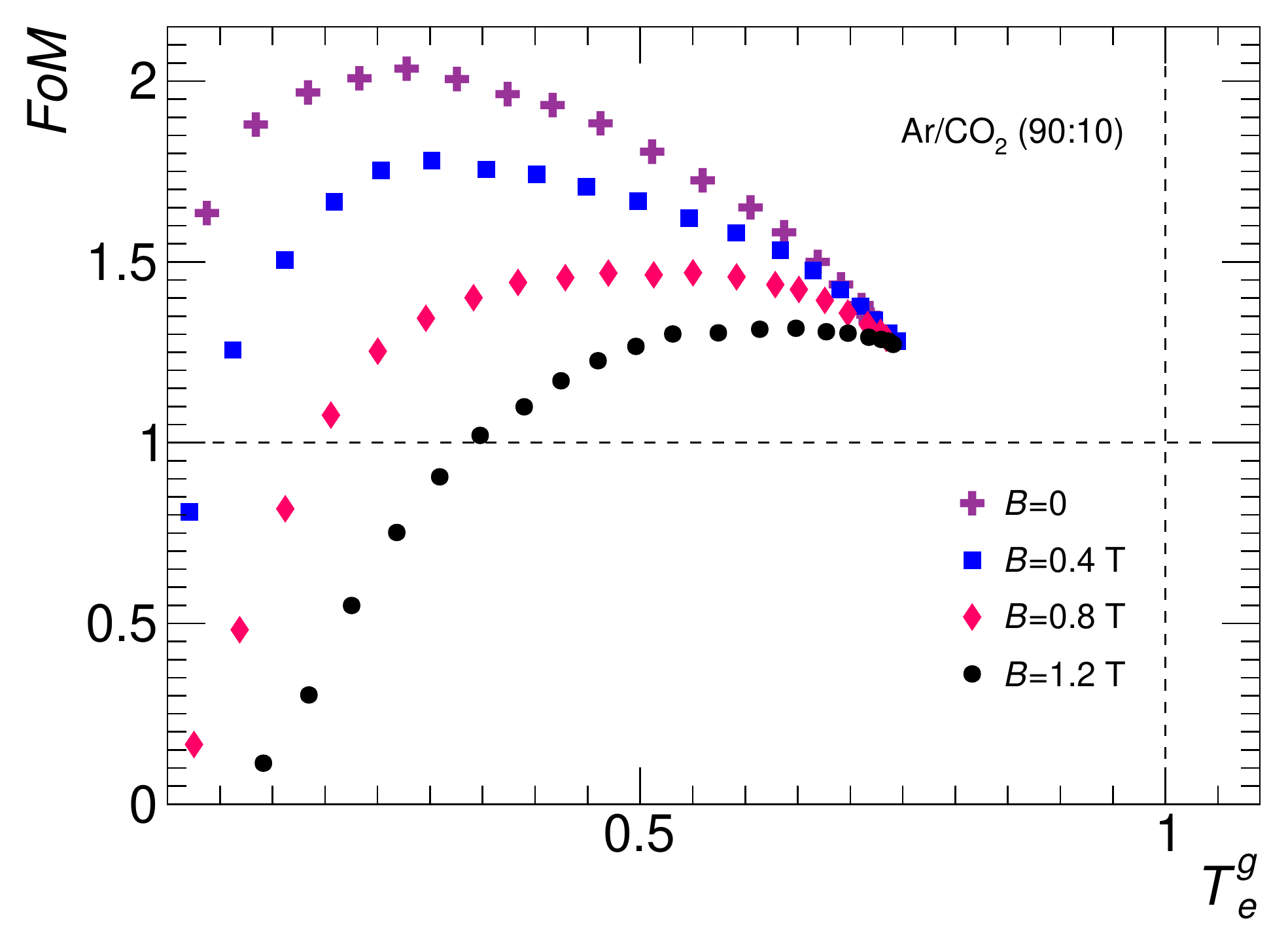}
\caption{\BPG performance in Ar/CO$_{2}$ (90:10) gas mixture at $\Ed$=320~V/cm, $\Et$=480~V/cm. Left: Transparency as a function of \dV. Right: \FoM vs. \Teg.}
\label{fig:fom_vs_teg_ArCO2_90_10}
\end{figure*}

\begin{figure*}[htb] 
\centering
\includegraphics*[width=0.49\textwidth]{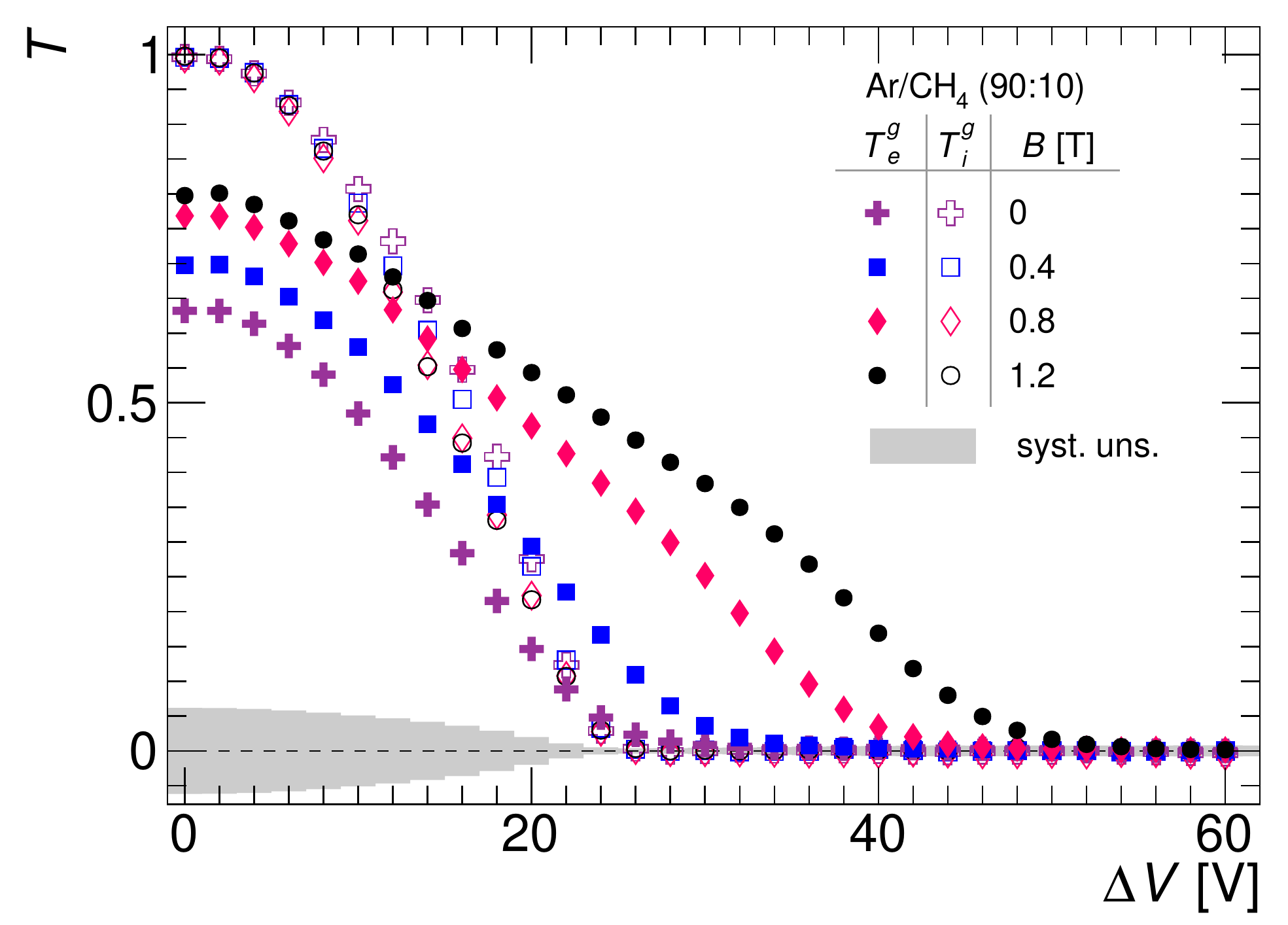}
\includegraphics*[width=0.49\textwidth]{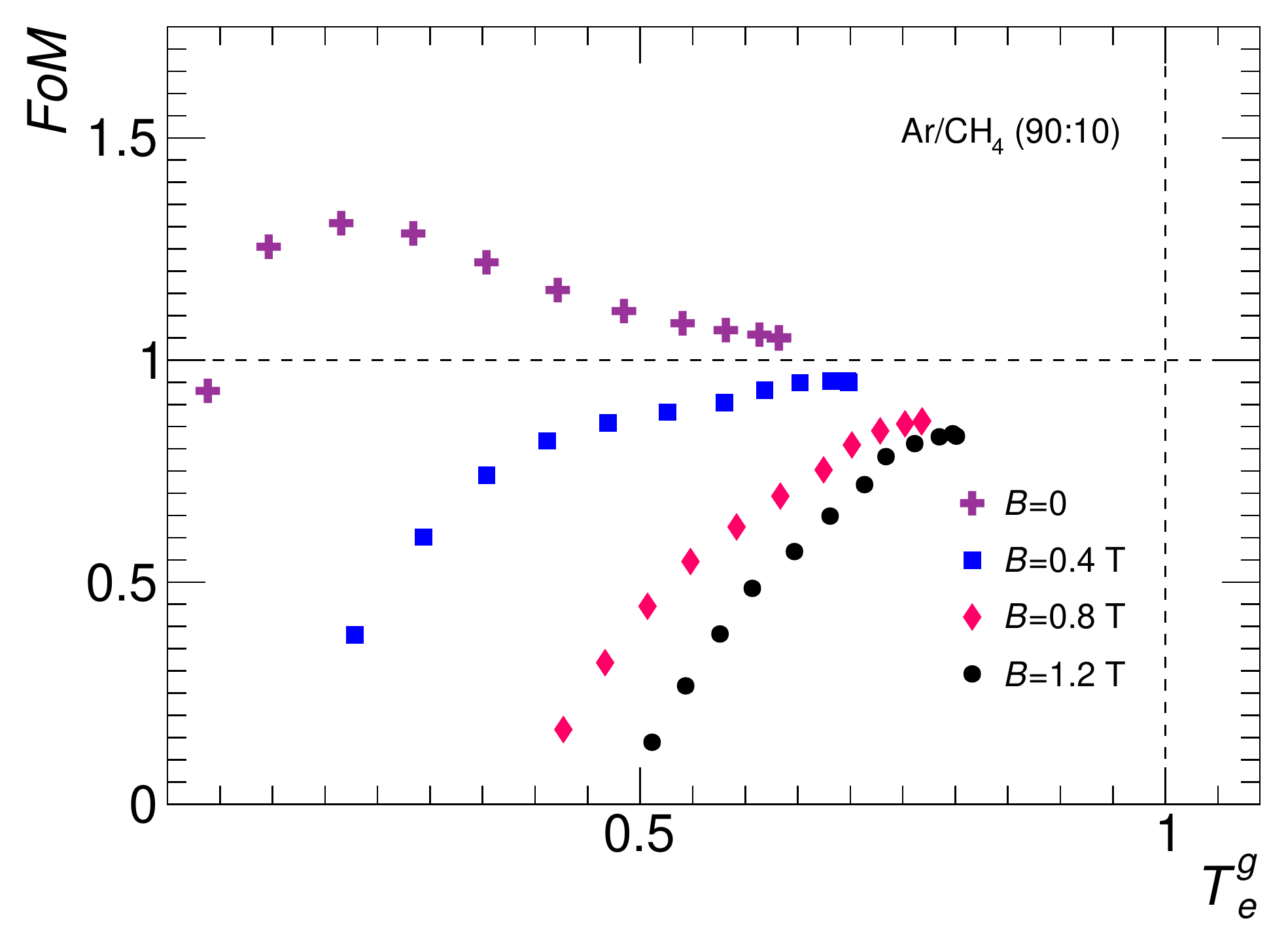}
\caption{\BPG performance in Ar/CH$_{4}$ (90:10) gas mixture at $\Ed$=320~V/cm, $\Et$=160~V/cm. Left: Transparency as a function of \dV. Right: \FoM vs. \Teg.}
\label{fig:fom_vs_teg_ArCH4_90_10_Et0p5}
\end{figure*}

\begin{figure*}[htb] 
\centering
\includegraphics*[width=0.49\textwidth]{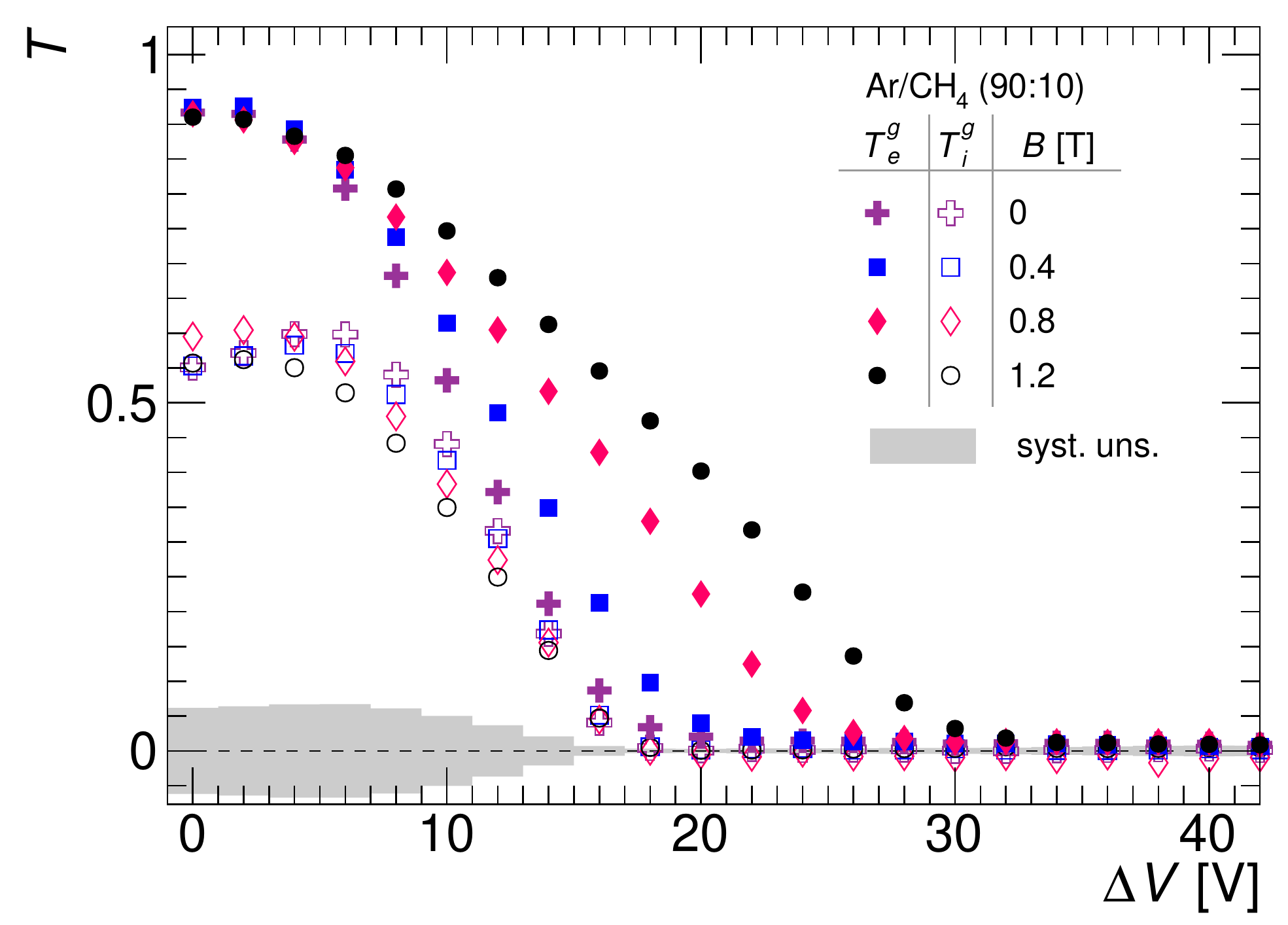}
\includegraphics*[width=0.49\textwidth]{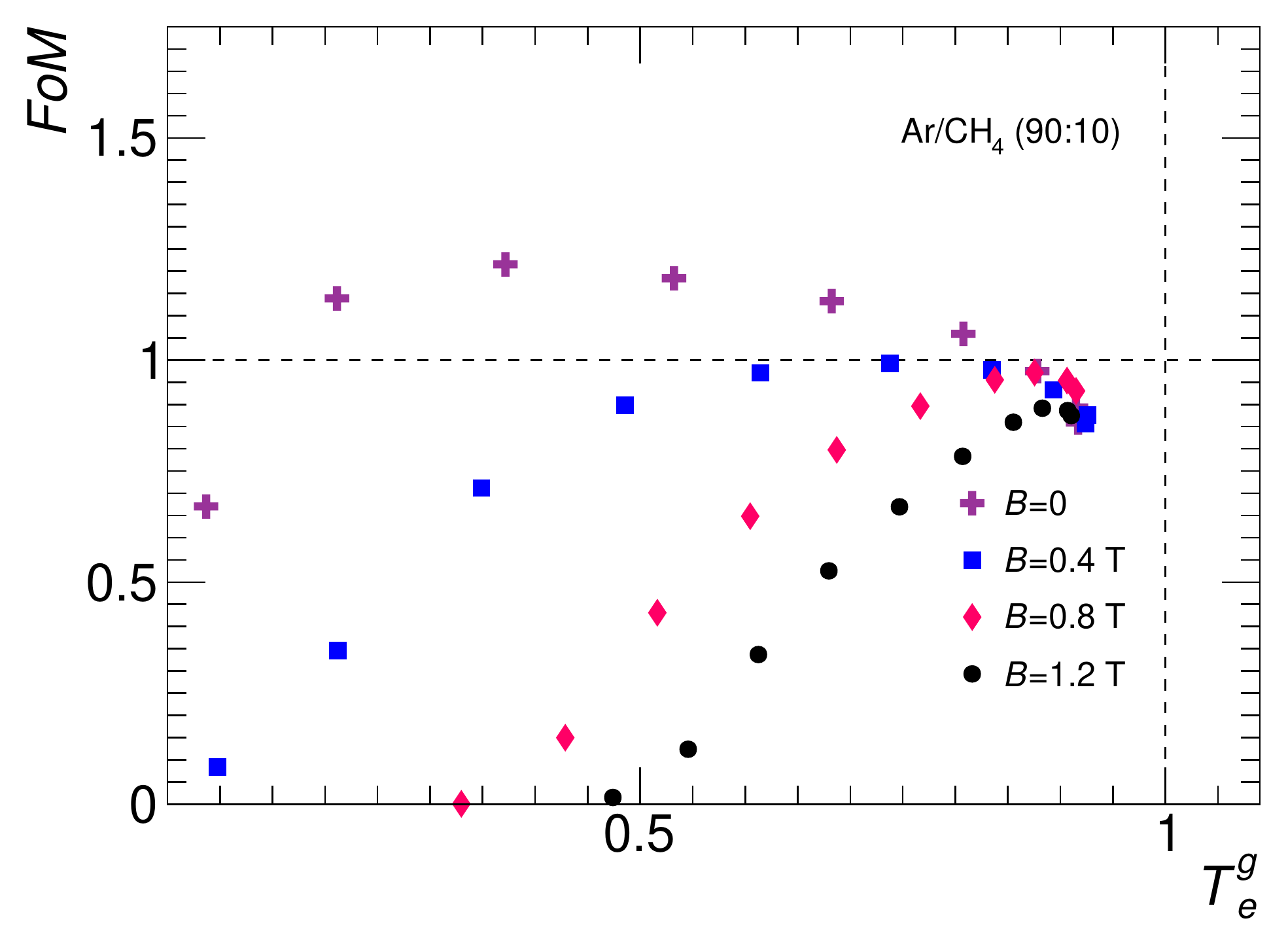}
\caption{\BPG performance in Ar/CH$_{4}$ (90:10) gas mixture at $\Ed$=140~V/cm, $\Et$=210~V/cm. Left: Transparency as a function of \dV. Right: \FoM vs. \Teg.}
\label{fig:fom_vs_teg_ArCH4_90_10_Ed140}
\end{figure*}

\bibliography{refs}


\end{document}